\documentclass[10pt]{article}
\usepackage{amsmath,amssymb}
\usepackage{mathtools}
\usepackage{color}
\usepackage[a4paper,text={16.8cm,22.4cm}]{geometry}
\usepackage{graphicx}
\usepackage{caption}
\usepackage{float}
\usepackage{subcaption}
\usepackage{appendix}
\usepackage{booktabs}
\usepackage{bm}
\usepackage{jheppub}
\usepackage{lmodern}
\usepackage{feynmp-auto}
\usepackage{appendix}
\usepackage{slashed}
\allowdisplaybreaks[1]

\title{\bfseries Partial Next-to-Next-to-Leading-Order Corrections to the $B \to \pi  $ Form Factors from Light-Cone Sum Rules}

\author[a]{Yong-Kang Huang,}
\emailAdd{huangyongkang@mail.nankai.edu.cn}

\author[,b,c,d]{Dong-Hao Li,}
\emailAdd{lidonghao@ihep.ac.cn}

\author[c,d]{Cai-Dian L\"u,}
\emailAdd{lucd@ihep.ac.cn}

\author[,a]{Bo-Xuan Shi,}
\emailAdd{shibx@mail.nankai.edu.cn}

\author[,a]{Hui-Xin Yu}
\emailAdd{yuhuixin@mail.nankai.edu.cn}

\affiliation[a]{
    School of Physics, Nankai University, No. 94 Weijin Road, Tianjin 300071, China}

 \affiliation[b]{
MOE Frontiers Science Center for Rare Isotopes, \\and School of Nuclear Science and Technology, Lanzhou University, Lanzhou 730000, China}

\affiliation[c]{
    Institute of High Energy Physics, CAS, P.O. Box 918(4), Beijing 100049,  China}

\affiliation[d]{
    School of Physics, University of Chinese Academy of Sciences, Beijing 100049, China}

\abstract{
By incorporating the available leading-power results at $\mathcal{O}(\alpha_s)$ and next-to-leading-power corrections at tree level, we compute the $\mathcal{O}(\alpha_s^2\beta_0)$ contribution to the $B\to\pi$ form factors in the large-recoil region using the light-cone sum rule approach with $B$-meson light-cone distribution amplitudes.
 We find that the QCD corrections at $\mathcal{O}(\alpha_s^2\beta_0)$ contribute approximately $+6.1\%$ compared to the tree-level result.
Combining the light-cone sum rule predictions in the large-recoil region, lattice QCD results in the small-recoil region, we perform a combined fit for the $B\to\pi$ form factors across the full kinematic range. Utilizing these form factors, we update the branching ratios, lepton-flavor-universality ratio $R_\pi$, forward-backward asymmetry $\mathcal{A}_{\rm FB}$, flat term  $\mathcal{F}_{\rm H}$ and polarization asymmetry $\mathcal{A}_{ \lambda_\ell}$  of $B\to\pi\mu\bar{\nu}_{\mu}$ and $B\to\pi\tau\bar{\nu}_{\tau}$ decays.
We also extract the Cabibbo-Kobayashi-Maskawa matrix element $|V_{ub}| = 3.73(14) \times 10^{-3}$.
We find that the resulting shifts are generally small and remain within the uncertainties of previous analyses.
}

\begin{document}

    \maketitle
    \newpage

\section{Introduction}

Since the establishment of the Standard Model, exploring new physics beyond the Standard Model has become the central goal for particle physicists. Besides the high-energy frontier in the direct searches for new physics signal, the meticulous scrutiny of the Standard Model plays a crucial role.
Heavy flavor physics, supported by a wealth of experimental data, has made a great contribution to precise testing of the Standard Model. One of the most important tasks is testing the unitarity of the Cabibbo-Kobayashi-Maskawa (CKM) matrix.

There exist various long-standing anomalies in the heavy flavor sector which serve as indirect avenues to explore new physics. One crucial anomaly arises from the tension between the $\left|V_{ub}\right|$ determined by the inclusive process $b \to u \ell^-\bar{\nu}_{\ell}$ and that through exclusive decays $\bar{B}^0\to\pi^+\ell^-\bar{\nu}_{\ell}$ \cite{HeavyFlavorAveragingGroup:2022wzx}.
Theoretical calculations of inclusive decays have reached high precision, thus the uncertainty in extraction of $\left|V_{ub}\right|$ from inclusive decays is mainly from the systematic error of  experiments and the unknown error of   quark hadron duality approximation.
In fact, experimental measurements of exclusive decays typically achieve better precision than those of inclusive decays. Thus the uncertainty of $\left|V_{ub}\right|$ determination in exclusive decays is mainly from the theoretical side.
Since the branching ratio for the $\bar{B}^0\to\pi^+\ell^-\bar{\nu}_{\ell}$ process is proportional to the transition form factors, the primary source of uncertainty in the extraction of $\left|V_{ub}\right|$ arises from theoretical calculation of the $B\to\pi$ transition form factors.

In the small recoil region, non-perturbative contributions dominate the $B\to\pi$ hadronic transition form factors, which have been investigated through lattice QCD (LQCD) simulations \cite{Colquhoun:2015mfa,Colquhoun:2022atw,Dalgic:2006dt,FermilabLattice:2015cdh,FermilabLattice:2015mwy,Flynn:2015mha} with high precision.
In the large-recoil region, although non-perturbative effects are also essential, the form factors can be analyzed using QCD-based methods in which short-distance contributions are treated perturbatively and long-distance dynamics are encoded in simpler hadronic objects such as light-cone distribution amplitudes (LCDAs).
These methods include light-cone sum rules (LCSRs) based on the {LCDAs} of $B$-meson \cite{Khodjamirian:2005ea,Khodjamirian:2006st,Faller:2008tr,DeFazio:2005dx,DeFazio:2007hw} or light mesons \cite{Duplancic:2008ix,Bharucha:2012wy,Rusov:2017chr,Cheng:2017sfk,Cheng:2025hxe},
transverse-momentum-dependent QCD factorization \cite{Li:2012nk,Li:2014xda,Cheng:2014fwa},
 QCD factorization \cite{Beneke:2000wa} and Soft-Collinear Effective Theory (SCET) \cite{Beneke:2002ni,Beneke:2002ph,Beneke:2003pa,Cui:2023jiw}.

At leading power in $\Lambda_{\rm QCD}/m_b$, the $B\to\pi$ form factors in the large-recoil region can be described in the QCD factorization/SCET framework as a sum of an A-type contribution and a B-type contribution \cite{Beneke:2005gs},
\begin{equation}
\label{QCD-SCET-FF}
f^i_{B\pi}(n\cdot p)=C_i^{(\mathrm A)}(n\cdot p)\,\xi_{\pi}(n\cdot p)
+\int d\tau\, C_i^{(\mathrm B)}(\tau,n\cdot p)\,\Xi_{\pi}(\tau,n\cdot p),
\end{equation}
where $i\in\{+,0,T\}$ labels the three $B\to\pi$ form factors, and $n\cdot p$ denotes the large light-cone projection of the pion momentum. The short-distance coefficients $C_i^{(\mathrm A)}$ and $C_i^{(\mathrm B)}$ are obtained from matching QCD onto SCET and have been computed up to one-loop and two-loop accuracy in the literature \cite{Bauer:2000yr,Beneke:2004rc,Bonciani:2008wf,Asatrian:2008uk,Beneke:2008ei,Bell:2010mg,Bell:2008ws,Becher:2004kk,Hill:2004if,Beneke:2005gs}.
The first term in Eq.~(\ref{QCD-SCET-FF}) involves the soft form factor $\xi_{\pi}$, which is a genuinely non-perturbative quantity. Owing to the presence of endpoint divergences, this contribution cannot be factorized further in a model-independent way and therefore has to be treated as an external non-perturbative input. The second term describes the hard-spectator-scattering contribution and involves the non-local form factor $\Xi_{\pi}(\tau,n\cdot p)$, where $\tau$ is the variable conjugate to the light-cone separation in the corresponding B-type SCET operator. After matching from SCET$_{\rm I}$ to SCET$_{\rm II}$, the latter can be factorized further into a convolution of light-cone distribution amplitudes with a hard-collinear jet function \cite{Beneke:2005gs}.

 The LCSR approach follows a strategy akin to traditional QCD sum rules.
Using the LCSR approach based on the LCDAs of light mesons, systematic studies have been carried out over the past years on heavy-to-light form factors, including: tree-level contributions up to twist-six, $\mathcal{O}(\alpha_s)$ corrections to the twist-two and twist-three contributions, and $\mathcal{O}(\alpha_s^2\beta_0)$ corrections to the twist-two contribution \cite{Rusov:2017chr,Khodjamirian:1997ub,Bagan:1997bp,Ball:2004ye,Khodjamirian:2011ub,Duplancic:2008ix,Bharucha:2012wy,Ball:2001fp}.
Alternatively, by considering the vacuum-to-$B$-meson correlation function, one can compute the two-point correlator in a kinematic region suitable for operator product expansion. Using perturbative QCD and the light-cone operator product expansion, the form factors can then be factorized into a hard scattering kernel convolved with the $B$-meson LCDA.
Loop-level calculations can be facilitated using the method of regions \cite{Beneke:1997zp}, which has been extensively applied in $B$-meson decays. An advantage of this LCSR approach is the emergence of a universal non-perturbative object, the $B$-meson LCDA, which appears in the expression for all $B$-meson decay form factors, including those $B$-to-heavy-meson transitions. This universality enables global fits to experimental data to constrain the $B$-meson LCDA. Employing this approach, systematic investigations of higher-order and subleading power corrections are feasible \cite{Shen:2016hyv,Shen:2021yhe,Gao:2019lta,Gubernari:2018wyi,Lu:2018cfc,Wang:2015vgv,Cui:2022zwm,Gao:2024vql,Khodjamirian:2023wol}.

To reduce the theoretical uncertainty in the extraction of $|V_{ub}|$, this paper focuses on the two-loop corrections to the $B\to\pi$ form factors within the LCSR approach based on the $B$-meson LCDA. In this framework, one considers a suitably defined light-cone correlation function with external momenta chosen in the hard-collinear region.
At leading power in the heavy-quark expansion, this correlation function factorizes into a convolution of the $B$-meson LCDA with perturbatively calculable functions associated with hard and hard-collinear fluctuations \cite{Wang:2015vgv}. The hard contribution can equivalently be identified with the matching coefficient obtained from matching QCD onto SCET$_{\rm I}$, for which the complete $\mathcal{O}(\alpha_s^2)$ result is already available in the literature.
In the present work, we concentrate on the two-loop corrections to the hard-collinear function. Rather than computing the full two-loop diagrams, we adopt the naive non-Abelianisation approximation, namely, we first evaluate the quark-loop contributions and then replace $N_f$ by $(-3/2)\beta_0$ \cite{Brodsky:1982gc,Broadhurst:1994se}.
Numerically, after incorporating the complete $\mathcal{O}(\alpha_s^2)$ correction to the hard function and the $\mathcal{O}(\alpha_s^2\beta_0)$ correction to the hard-collinear function, we find that the one-loop corrections amount to approximately $-30\%$ relative to the tree-level result, while the partial NNLO corrections enhance the form factors by about $+6.1\%$.
The $\mathcal{O}(\alpha_s^2\beta_0)$ results obtained in this work can also serve as a starting point for future investigations of renormalon effects in the $B\to\pi$ form factors.

The structure of the paper is outlined as follows:
In Section 2,  we construct the light-cone sum rule framework for the   $B\to\pi$ transition form factors and calculate the corresponding correlation function at the next-to-next-to-leading order (NNLO) in the large-$\beta_0$ limit.
In Section 3, we present our numerical results, phenomenological analysis and the extraction of $|V_{ub}|$.
We conclude in Section 4.
The technical details are presented in the Appendices, including soft subtraction, factorization-scale independence as consistency checks of QCD factorization, and the validity of the Wandzura–Wilczek relation between the two $B$-meson distribution amplitudes $\phi_{B}^+(\omega)$ and $\phi_B^-(\omega)$.

\section{\texorpdfstring{$B \to \pi$}{B to pi } transition form factors calculated in LCSR}

The hadronic matrix element for the $B \to \pi$ transition is parameterized by two form factors,
\begin{align}
    \langle\pi(p)\vert\bar{u}\gamma_{\mu}b\vert\bar{B}(p_B)\rangle=f^{+}_{B\pi}(q^2)\left[ 2p_{\mu}+\left( 1-\frac{m^2_B-m_\pi^2}{q^2}\right)q_{\mu}\right]+f_{B\pi}^{0}(q^2)\frac{m_B^2-m_\pi^2}{q^2}q_{\mu},
\end{align}
where the $B$-meson and the pion carry momenta $p_B^\mu$ and $p^\mu$, respectively.

The momentum transfer $q^\mu$ is defined as $q^\mu \equiv p_B^\mu - p^\mu = m_B v^\mu - p^\mu$.

The main objective of this work is to provide predictions for these two form factors in the large-recoil region.
To construct the light-cone sum rules for $f^{+}_{B\pi}(q^2)$ and $f^{0}_{B\pi}(q^2)$, we start with the correlation function:
\begin{align}
        \Pi_\mu(n\cdot p,\bar{n}\cdot p )&=\int d^4x~ e^{i p\cdot x} \langle 0 \vert T\left\{\bar{d}(x)\slashed{n}\gamma_5u(x),\bar{u}(0)\gamma_\mu b(0)\right\}\vert\bar{B}(p+q)\rangle\notag \\
        &=\Pi_n(n\cdot p,\bar{n}\cdot p )n_\mu + \Pi_{\bar{n}}(n \cdot p,\bar{n}\cdot p )\bar{n}_{\mu}
        \label{eq:define the correlation function},
\end{align}
which involves a pion interpolating current $\bar{d}(x)\slashed{n}\gamma_5 u(x)$ and a weak current $\bar{u}(0)\gamma_\mu b(0)$.

To formulate the sum rule in a way consistent with the standard SCET power counting, we introduce two light-like vectors $n^\mu$ and $\bar n^\mu$ satisfying $n^2=\bar n^2=0,~n\cdot \bar n=2,$ and decompose the momentum carried by the interpolating current as
\begin{align}
p^\mu=\frac{n\cdot p}{2}\,\bar n^\mu+\frac{\bar n\cdot p}{2}\,n^\mu+p_\perp^\mu .
\end{align}
We work in a frame with $p_\perp^\mu=0$ and adopt the large-recoil scaling
\begin{align}
n \cdot p \simeq \frac{m_B^2-q^2}{m_B}\sim {\cal O}(m_b),
\qquad
\bar n\cdot p <0,\qquad |\bar n\cdot p|\sim {\cal O}(\Lambda_{\rm QCD}) .
\label{power_counting_revised}
\end{align}
It then follows that the external momenta scale as $
|p^2|\sim {\cal O}(m_b\Lambda_{\rm QCD})$
which is the characteristic hard-collinear virtuality in the SCET sum-rule framework \cite{DeFazio:2007hw}.
In this setup, the momentum transfer is kept in the large-recoil region, such that $n\cdot p$ remains of order $m_b$ and $q^2$ stays away from the zero-recoil endpoint \cite{Khodjamirian:2006st}. The dispersive analysis is performed in the variable $\bar n\cdot p$ at fixed $n\cdot p$. For such momentum configurations, the light-cone correlation function admits an operator-product expansion: hard fluctuations with virtuality of order $m_b^2$ are integrated out into short-distance coefficient functions, while the perturbative dynamics of the correlator is governed by hard-collinear modes with virtuality of order $m_b\Lambda_{\rm QCD}$, and the nonperturbative effects are encoded in the $B$-meson LCDAs \cite{Wang:2015vgv},
\begin{align}
\Pi_{n,\bar{n}}=\tilde{f}_B(\mu)m_B\sum_{k=\pm}C^{(k)}_{n,\bar{n}}(n\cdot p,\mu)\int_0^\infty \frac{d\omega}{\omega-\bar{n}\cdot p}J^{(k)}_{n,\bar{n}}\left(\frac{\mu^2}{n\cdot p~ \omega},\frac{\omega}{\bar{n}\cdot p}\right)\phi^{(k)}_B(\omega,\mu).
\label{eq:fac formula2}
\end{align}

By applying the dispersion relation, the partonic representation of the correlation function can be written as
\begin{align}
\Pi_{n,\bar{n}}(n\cdot p,\bar{n}\cdot p)
= \frac{1}{\pi} \int_0^\infty \frac{d\omega'}{\omega'-\bar{n}\cdot p-i0}
\,{\rm Im}_{\omega'}\Pi_{n,\bar{n}}(n\cdot p,\omega') .
\label{eq:partonic_dispersion}
\end{align}
The hadronic representation of the same correlation function is given by
\begin{equation}
\begin{aligned}
\Pi_\mu^{\rm had}(n\cdot p,\bar{n}\cdot p)
=&~\frac{f_\pi m_B}{2(m_\pi^2/n\cdot p-\bar{n}\cdot p)}
\left\{
\left[ \frac{n\cdot p}{m_B} f_{B\pi}^{+}(q^2)+f_{B\pi}^{0}(q^2)\right] \bar{n}_\mu
\right.\\
&\left.
+~\frac{m_B}{n\cdot p-m_B}
\left[ \frac{n\cdot p}{m_B} f_{B\pi}^{+}(q^2)-f_{B\pi}^{0}(q^2) \right]n_\mu
\right\}\\
&+\int_{\omega_s}^\infty\frac{d\omega}{\omega-\bar{n}\cdot p-i0}
\left[ \rho_{\bar{n}}(n\cdot p,\omega)\bar{n}_\mu +\rho_n(n\cdot p,\omega)n_\mu \right].
\end{aligned}
\end{equation}
Here, $f_\pi$ is defined through
$\langle 0 \vert \bar{u} ~ \slashed{n} \gamma_5 ~ d\vert \pi(p)\rangle = i (n\cdot p) f_\pi .$

The parameter $\omega_s$ denotes the effective continuum threshold in the dispersion relation with respect to the variable $\bar n\cdot p$. Since $p^2=(n\cdot p)(\bar n\cdot p)$
the corresponding threshold in the conventional $p^2$ channel is
$
s_0=(n\cdot p)\,\omega_s
$
, the threshold $s_0$ scales with the large light-cone component $n\cdot p$ and varies with the recoil kinematics $q^2$.

\begin{figure}[tb]
        \centering
        \includegraphics[scale=0.35]{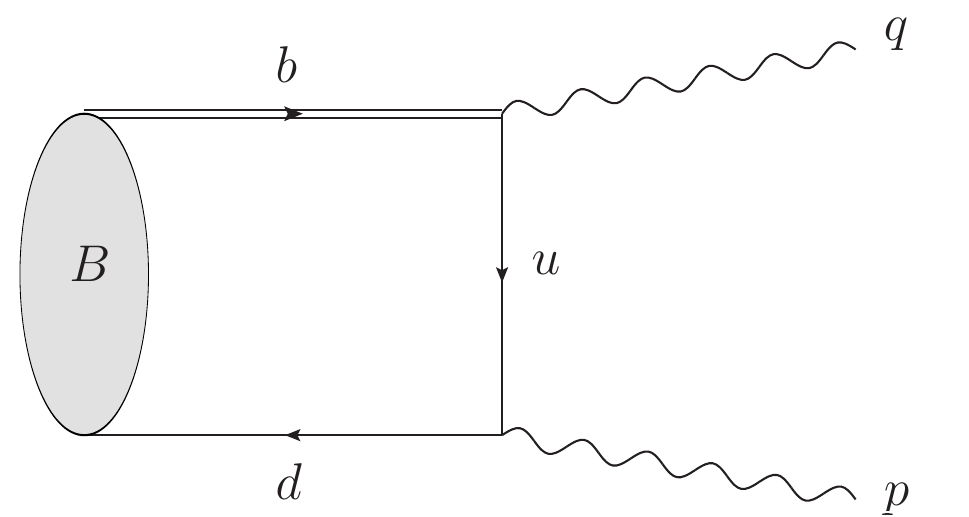}
        \caption{Diagrammatic representation of the correlation function $\Pi_\mu(n\cdot p, \bar{n}\cdot p)$ at tree level.}
        \label{fig:Tree level}
    \end{figure}

By invoking quark-hadron duality above the effective threshold $\omega_s$, the hadronic spectral densities are approximated by the corresponding QCD spectral functions,
\begin{align}
\int_{\omega_s}^\infty\frac{d\omega'}{\omega'-\bar{n}\cdot p -i0}\rho_{n,\bar{n}} (n \cdot p, \omega')
=\frac{1}{\pi}\int_{\omega_s}^\infty \frac{d\omega'}{\omega'-\bar{n}\cdot p-i0}
~{\rm Im}_{\omega'}\Pi_{n,\bar{n}}(n\cdot p,\omega') .
\label{eq:duality_revised}
\end{align}
After performing the Borel transformation with respect to the variable $\bar n\cdot p \to \omega_M$, one obtains the general sum-rule relation
\begin{align}
& f_\pi m_B ~ {\rm exp}\left[-\frac{m_\pi^2}{n\cdot p~ \omega_M}\right]
\left\{\frac{n\cdot p}{m_B}f_{B\pi}^+(q^2),\,f_{B\pi}^0(q^2)\right\}
\nonumber \\
&~=\int_0^{\omega_s}d \omega' ~{\rm exp}\left[-\frac{\omega'}{\omega_M}\right]
\frac{1}{\pi}\left[
{\rm Im}_{\omega'}\left(\Pi_{\bar{n}} (n\cdot p,\omega')
\pm \frac{n\cdot p-m_B}{m_B}\Pi_n(n\cdot p,\omega')\right)\right] ,
\label{eq:9}
\end{align}
{ where the two entries in braces on the left-hand side are understood to correspond in order to the upper and lower signs on the right-hand side. The upper sign refers to the sum rule for $\frac{n\cdot p}{m_B}f_{B\pi}^+(q^2)$, while the lower sign refers to that for $f_{B\pi}^0(q^2)$.
}
The last expression can be further simplified at tree level by evaluating the diagram shown in Fig.\ref{fig:Tree level} in the heavy-quark and large-recoil limit. This yields
\begin{align}
 f_\pi m_B ~ {\rm exp}\left[-\frac{m_\pi^2}{n\cdot p~ \omega_M}\right]
\left\{\frac{n\cdot p}{m_B}f_{B\pi}^+(q^2),\,f_{B\pi}^0(q^2)\right\}
= m_B \tilde{f}_B \int_0^{\omega_s}d \omega' ~{\rm exp}\left[-\frac{\omega'}{\omega_M}\right]
 \phi_B^- (\omega')
+\mathcal{O}(\alpha_s).
\label{eq:tree_sum_rule}
\end{align}
{At tree level, the correlation-function component $\Pi_n$ vanishes, $\Pi_n^{(0)}=0$. Consequently, the term proportional to $\pm (n\cdot p-m_B)\Pi_n/m_B$ in Eq.~\eqref{eq:9} does not contribute, and the two sum rules therefore coincide at this order, giving the same right-hand side in Eq.~\eqref{eq:tree_sum_rule}.
}
For completeness, we recall that the two-particle $B$-meson LCDAs $\phi_B^\pm$ are defined by
\begin{align}
\langle0 \vert \bar{d} (\tau \bar{n})\left[\tau\bar{n},0\right]
(\slashed{n}\gamma_5)b(0)\vert\bar{B}(p_B)\rangle
&=\frac{i\tilde{f}_B(\mu)m_B}{4}{\tilde{\phi}_B^-(\tau)}, \notag \\
\langle0 \vert \bar{d} (\tau \bar{n})\left[\tau\bar{n},0\right]
(\slashed{\bar{n}}\gamma_5)b(0)\vert\bar{B}(p_B)\rangle
&=\frac{i\tilde{f}_B(\mu)m_B}{4}{\tilde{\phi}_B^+(\tau)},
\end{align}
with the Fourier transformations
\begin{align}
\phi_B^{\pm}(\omega) = \int_{-\infty}^{\infty} \frac{d\tau}{2\pi} e^{i\omega \tau} \tilde{\phi}_B^{\pm} (\tau - i0),
\end{align}
where $\tilde f_B$ is the $B$-meson decay constant defined in HQET. For the present correlation function, only $\phi_B^-(\omega)$ enters the tree-level sum rule in Eq.~\eqref{eq:tree_sum_rule}.

{Beyond the leading-power tree-level result, the corrections to the correlation function can be organized into perturbative and power-suppressed contributions. The former arise from radiative corrections, while the latter are suppressed by $\Lambda_{\rm QCD}/m_b$ or $\Lambda_{\rm QCD}/(n\cdot p)$. At next-to-leading power, the known tree-level contributions arise from several sources, including the power expansion of the hard-collinear quark propagator, the subleading heavy-to-light effective current, and higher-twist $B$-meson LCDAs~\cite{Cui:2022zwm}. These corrections will be included in our numerical analysis.
In the following, we focus on the perturbative corrections to the leading-power correlation function.
}

\subsection{Factorization formulae for the correlation function beyond leading order}

At leading power in the heavy-quark expansion, the correlation function $\Pi_{n,\bar{n}}(n\cdot p, \bar{n}\cdot p)$ can be written in the standard factorized form \cite{Wang:2015vgv},
\begin{align}
\Pi_{n,\bar{n}}=\tilde{f}_B(\mu)m_B\sum_{k=\pm}C^{(k)}_{n,\bar{n}}(n\cdot p,\mu)\int_0^\infty \frac{d\omega}{\omega-\bar{n}\cdot p}J^{(k)}_{n,\bar{n}}\left(\frac{\mu^2}{n\cdot p~ \omega},\frac{\omega}{\bar{n}\cdot p}\right)\phi^{(k)}_B(\omega,\mu),
\label{eq:fac formula}
\end{align}
The hard coefficient functions $C^{\pm}_{n,\bar{n}}$ can be obtained from the matching coefficients of the QCD current $\bar{q}\gamma^\mu b(0)$ in SCET\textsubscript{I},
\begin{align}
    (\bar{q}\gamma^\mu b)(0)  = &\int ds \, \sum_i \tilde{C}_i^{(A0)}(s) (\bar{\xi}W_c)(s \bar{n}) \Gamma^{A0}_{i} h_v(0) \nonumber \\
    & + \int ds_1 ds_2 \sum_j \tilde{C}_j^{(B1)}(s_1,s_2) \frac{1}{m_b} (\bar{\xi}W_c)(s_1 \bar{n}) \left [ W_c^\dagger i \slashed D_{\perp c}W_c \right ](s_2 \bar{n}) \Gamma^{B1}_{j} h_v(0)
    + \cdots
    \label{eq:scet1 operator}
\end{align}
where $\Gamma^{A0}_{i}\in \{ \gamma^\mu, v^\mu, \bar{n}^\mu \}$, $\Gamma^{B1}_{j}\in \{  v^\mu, \bar{n}^\mu \}$ and the ellipsis denotes other operators or Dirac structures that do not contribute to the form factors $f_{B\pi}^{+,0}(q^2)$ \cite{Beneke:2005gs}.
Here $s$, $s_1$ and $s_2$ denote the light-cone position variables of the non-local SCET operators. Following the conventions of \cite{Beneke:2004rc,Beneke:2005gs}, it is convenient to introduce the dimensionless variables $\hat s_i=s_i m_b/(n \cdot v)$, and the momentum-space hard coefficients are defined by Fourier transformation with respect to $\hat s_i$,
\begin{align}
&C_i^{(A0)}(n\cdot p)=\int d\hat{s}~ e^{i \hat{s} \, n\cdot p/m_b} ~\tilde{C}_i^{(A0)}(s),
\nonumber
\\
&C_j^{(B1)}(n\cdot p_1,n\cdot p_2)=\int d\hat{s}_1 \,d\hat{s}_2~ e^{i \sum_i \hat{s}_i \, n\cdot p_i/m_b} ~\tilde{C}_j^{(B1)}(s_1,s_2),
\end{align}
where the A-type coefficient functions depend on the large light-cone momentum $n\cdot p$, while for the B-type three-body operators the momentum-space coefficients depend on the large light-cone momenta carried by the collinear quark and gluon fields. These variables can be reorganized into the total large light-cone momentum $n\cdot p$ and a momentum fraction $u \in [0,1]$. Therefore, there is in principle also a convolution integral over $u$ in the B-type contribution. In the factorization formula this additional convolution is trivial and has been omitted for brevity.
Inserting Eq.\eqref{eq:fac formula} into Eq.\eqref{eq:define the correlation function} and comparing with Eq.\eqref{eq:scet1 operator} yields the relation
\begin{align}
    C^{(+)}_n &=\frac{1}{2} C^{(B1)}_1,\quad C^{(+)}_{\bar{n}} =\frac{1}{2} C^{(B1)}_1+ C^{(B1)}_2 ,  \nonumber \\
    C^{(-)}_n &= \frac{1}{2} C^{(A0)}_2,\quad C^{(-)}_{\bar{n}} = C^{(A0)}_1+\frac{1}{2} C^{(A0)}_2+ C^{(A0)}_3,
\end{align}
The hard coefficients $C_i^{(A0)}$ have been calculated at $\mathcal{O}(\alpha_s^2)$ in Ref.\cite{Beneke:2008ei} and $C_i^{(B1)}$ have been determined at $\mathcal{O}(\alpha_s)$ in Refs.\cite{Beneke:2004rc,Beneke:2005gs}.
Adopting the factorization scale $\mu $ of order $\sqrt{m_b\Lambda_{\text{QCD}}}$, we then perform the resummation of the large logarithms in the hard coefficients and the HQET decay constant $\tilde{f}_B$ to derive their scale dependence,
\begin{align}
        C^{(+)}_{n,\bar{n}}(n\cdot p ,\mu)&=U_{1} (n\cdot p, \mu_{ h1},\mu)C^{(+)}_{n,\bar{n}}(n\cdot p, \mu_{ h1}), \quad
        \tilde{f}_B(\mu)=U_2(\mu_{ h2},\mu)\tilde{f}_B(\mu_{ h2}),
  \notag\\
        C^{(-)}_{n,\bar{n}}(n\cdot p ,\mu)&=U_{3} (n\cdot p, \mu_{ h1},\mu)C^{(-)}_{n,\bar{n}}(n\cdot p, \mu_{ h1}),
  \label{eq:resum of hard func}
\end{align}
where the evolution kernel $U_2(\mu_{h2},\mu)$ and $U_{3} (n\cdot p, \mu_{ h1},\mu)$ at next-to-next-to-leading logarithm (NNLL) accuracy are collected in our previous work \cite{Cui:2023yuq}, and the evolution kernel of hard function $C^{(+)}_{n,\bar{n}}(n\cdot p ,\mu)$ at leading logarithm   accuracy can be obtained from Ref.\cite{Beneke:2005gs}.

To derive the jet function $J_{n,\bar n}^{\pm}$ at $\mathcal{O}(\alpha_s^2 \beta_0)$ within the factorized representation of the correlation function $\Pi_{n,\bar n}$, we introduce the {\it partonic} analogue of the vacuum-to-$B$-meson correlation function by replacing the external hadronic state with the on-shell partonic state $|b(p_B-k)\,\bar d(k)\rangle$. The correlator can then be expanded as
\begin{align}
    \Pi_{\mu,b \bar{d}}
    =&\Pi_{\mu,b \bar{d}}^{(0)}+\Pi_{\mu,b \bar{d}}^{(1)}+\Pi_{\mu,b \bar{d}}^{(2)}+\cdots
    =\Phi_{b\bar{d}}\otimes T\notag\\
    =&\Phi_{b\bar{d}}^{(0)}\otimes T^{(0)}
    +\left[\Phi_{b\bar{d}}^{(0)}\otimes T^{(1)}+\Phi_{b\bar{d}}^{(1)}\otimes T^{(0)}\right]\notag\\
    &+\left[\Phi_{b\bar{d}}^{(0)}\otimes T^{(2)}+\Phi_{b\bar{d}}^{(1)}\otimes T^{(1)}+\Phi_{b\bar{d}}^{(2)}\otimes T^{(0)}\right]+\cdots\, ,
\label{eq:FAEQ}
\end{align}
following Refs.~\cite{Descotes-Genon:2002crx,Wang:2015vgv}.
The partonic distribution amplitude of the $B$ meson is defined by
\begin{align}
    \Phi_{b \bar d}(\omega',k)
    =
    \int \frac{d \tau}{2 \pi}\, e^{i \omega' \tau}
    \langle 0|
    \bar d_{\beta}(\tau \bar n)[\tau \bar n, 0] b_{\alpha}(0)
    |b(p_B-k)\bar d(k)\rangle ,
\end{align}
where the dependence on the external partonic momentum $k$ will be suppressed whenever no confusion can arise. At tree level, one obtains
\begin{align}
    \Phi_{b \bar d}^{(0)}(\omega',k)
    =
    \delta(\bar n \cdot k-\omega')\,
    \bar d_{\beta}(k)\, b_{\alpha}(p_B-k).
\end{align}
Therefore, $\omega'$ is the convolution variable in the matching formula and should not, in general, be identified with $\bar n\cdot k$, although the two coincide at tree level.
With the help of the method of regions, the hard function $C$ and the jet function $J$ at $\mathcal{O}(\alpha_s)$ can be straightforwardly obtained from the matching condition:
\begin{align}
        \Phi_{b\bar{d}}^{(0)}\otimes T^{(1)}&=C^{(1)}\cdot J^{(0)} \otimes \Phi_{b\bar{d}}+C^{(0)} \cdot J^{(1)} \otimes \Phi_{b\bar{d}} \nonumber\\
&=\Pi_{\mu,b\bar{d}}^{(1),h}+\Pi_{\mu,b\bar{d}}^{(1),hc}+\Pi_{\mu,b\bar{d}}^{(1),s}-\Phi_{b\bar{d}}^{(1)}\otimes T^{(0)} .
\end{align}
In the above equations, the numerical superscripts $(0),(1)$ denote the order in the $\alpha_s$ expansion, the superscripts $h$, $hc$, and $s$ refer to contributions from the hard, hard-collinear, and soft regions, respectively and $\Phi_{b\bar{d}}$ is the partonic distribution amplitude of the $B$-meson,
\begin{align}
    \Phi_{b \bar{d}}\left(\omega^{\prime}\right)=\int \frac{d \tau}{2 \pi} e^{i \omega^{\prime} \tau}\langle 0| \bar{d}_{\beta}(\tau \bar{n})[\tau \bar{n}, 0] b_{\alpha}(0)\left|b\left(p_{B}-k\right) \bar{d}(k)\right\rangle.
\end{align}
The soft subtraction term $\Phi_{b\bar{d}}^{(1)}\otimes T^{(0)}$ precisely cancels the soft region contribution $\Pi_{\mu,b\bar{d}}^{(1),s}$, and the collinear region does not contribute at leading power. Consequently, the one-loop jet function $J^{(1)}$ can be extracted from the hard-collinear region contribution $\Pi_{\mu,b\bar{d}}^{(1),hc}$, as shown explicitly in Ref.~\cite{Wang:2015vgv}.
In the remainder of this section, we will compute $\Pi_{\mu,b\bar{d}}^{(2),hc}$ to extract the jet function $J$ at $\mathcal{O}(\alpha_s^2\beta_0)$ accuracy, noting that the term $\Phi_{b\bar{d}}^{(1)}\otimes T^{(1)}$ does not contribute to $\Pi_{\mu,b\bar{d}}^{(2),hc}$ at large-$\beta_0$ limit. The soft subtraction and the factorization-scale independence at $\mathcal{O}(\alpha_s^2\beta_0)$ accuracy, as consistency checks of the factorized expression, are discussed in the Appendix \ref{app-softsub}.

\subsection{Two-loop QCD corrections to the correlator \texorpdfstring{$\Pi_{\mu}$}{Pi} in the large-\texorpdfstring{$\beta_0 $}{beta0} approximation}

Computing the full two-loop radiative corrections to the correlation function is highly challenging.
In this work, we restrict ourselves to the $\beta_0$-enhanced part of the two-loop hard-collinear correction.
This contribution is obtained by evaluating the fermion-loop diagrams and then applying the naive non-Abelianization replacement
$N_f \to -3\beta_0/2$.
This procedure was first proposed by Brodsky, Lepage, and Mackenzie as a renormalization scale-setting method~\cite{Brodsky:1982gc}, in analogy with QED, where all effects from the running coupling are associated with photon vacuum polarization. They suggested absorbing the $\beta_0$-dependent contribution (from fermion-loop insertions) into an effective strong coupling constant~\cite{Beneke:1994qe}.
This approximation has been used as a heuristic diagnostic tool in several heavy-quark observables \cite{Broadhurst:1994se}, although its reliability is observable-dependent and can be assessed a posteriori. It has also been employed to compute the $\mathcal{O}(\alpha_s^2\beta_0)$ correction to the $B \to \pi$ form factors in the pion-LCDA LCSR framework~\cite{Bharucha:2012wy}.

 \begin{figure}[thb]
        \centering
        \begin{subfigure}[t]{0.2\textwidth}
            \includegraphics[width=\textwidth]{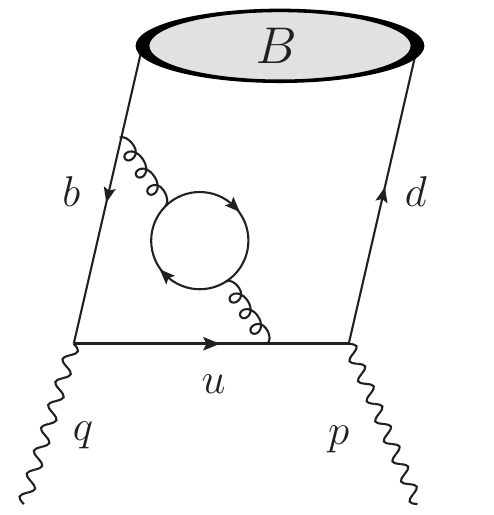}
            \caption{Weak vertex correction}
            \label{fig:weak}
        \end{subfigure}
        \hspace{0.05\textwidth}
        \begin{subfigure}[t]{0.2\textwidth}
            \includegraphics[width=\textwidth]{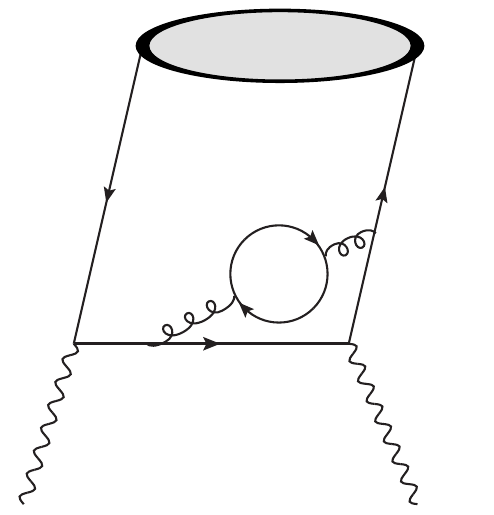}
            \caption{Pion vertex correction}
            \label{fig:pion}
        \end{subfigure}
        \hspace{0.05\textwidth}
        \begin{subfigure}[t]{0.2\textwidth}
            \includegraphics[width=\textwidth]{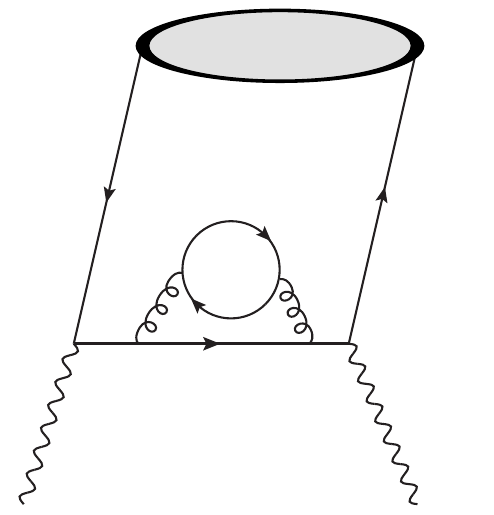}
            \caption{Self-energy correction}
            \label{fig:self}
        \end{subfigure}
        \hspace{0.05\textwidth}
        \begin{subfigure}[t]{0.2\textwidth}
            \includegraphics[width=\textwidth]{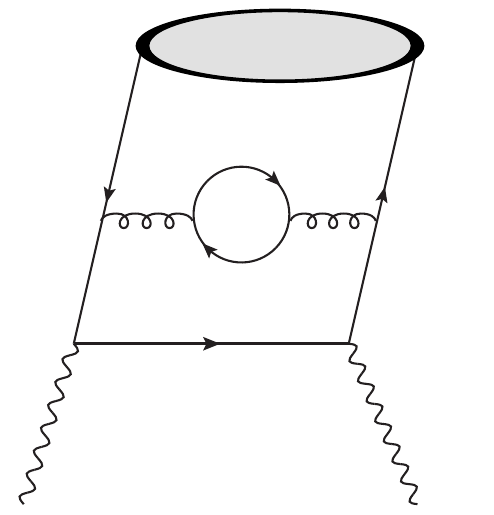}
            \caption{Box diagram correction}
            \label{fig:box}
        \end{subfigure}
        \captionsetup{labelfont=bf}
        \caption{Fermion-loop contributions to the correlation function $\Pi_\mu(n\cdot p,\bar{n}\cdot p)$ at $\mathcal{O}(\beta_0 \alpha_s^2)$.}
        \label{fig:two_loop feynnman diagrams}
    \end{figure}

In the large-$\beta_0$ limit, only the four fermion-loop diagrams shown in Fig.\ref{fig:two_loop feynnman diagrams} and their corresponding counterterms are required to compute $\Pi_{\mu,b\bar{d}}^{(2),hc}$.
These four two-loop diagrams share a common feature: the fermion-loop subdiagram is universal and can be absorbed into an effective vacuum-polarization insertion in the gluon propagator. For convenience, we introduce the effective dressed gluon propagator
\begin{align}
    \mathcal{D}^{ab}_{\mu\nu}(\ell)\equiv
    \frac{8i(-1)^{-\epsilon}}{(4\pi)^{2-\epsilon}}
    \frac{\Gamma(\epsilon)\Gamma^2(2-\epsilon)}{\Gamma(4-2\epsilon)}
    \,\delta^{ab} g_s^2 T_F N_f\,
    \frac{1}{(\ell^2)^{1+\epsilon}}
    \left(g_{\mu\nu}-\frac{\ell_\mu \ell_\nu}{\ell^2}\right),
    \label{eq:dressed-gluon-propagator}
\end{align}
such that the replacement of the tree-level gluon propagator can be written as
\begin{align}
    \frac{-ig_{\mu\nu}\delta^{ab}}{\ell^2}\;\longrightarrow\; \mathcal{D}^{ab}_{\mu\nu}(\ell).
    \label{eq:quark loop to gloun}
\end{align}
Here $g_s^2=4\pi\alpha_s$, $T_F=1/2$, and $N_f$ denotes the number of light-quark flavors. With this notation, the $\mathcal{O}(\alpha_s^2\beta_0)$ calculation can be viewed as the corresponding NLO diagrams with the gluon propagator replaced by the universal building block $\mathcal{D}^{ab}_{\mu\nu}(\ell)$. This representation allows us to analyze this pNNLO contribution in close analogy with the NLO hard-collinear calculation.

We first compute the contribution to $\Pi_\mu^{(2)}$ arising from the QCD correction to the weak vertex diagram in dimensional regularization ($D = 4 - 2\epsilon$), depicted in Fig.\ref{fig:weak},
\begin{align}
        \Pi_\mu^{(2a)}(p_B,q) =&{\mu^{4\epsilon} \frac{g_s^4 C_FT_FN_f}{(p-k)^2}\frac{-8i}{(4\pi)^{2-\epsilon}} \frac{\Gamma(\epsilon)\Gamma^2(2-\epsilon)}{\Gamma(4-2\epsilon)}}\notag\\
        &\times\int\frac{d^D\ell}{(2\pi)^D} \frac{\ell^2g_{\rho\rho'}-\ell_\rho \ell_{\rho'}}{\left[(p-k+\ell)^2+i0\right]\left[(p_B-k+\ell)^2-m_b^2+i0\right] (\ell^2+i0)^2 (-\ell^2)^\epsilon}\notag\\
        &\qquad\times\bar{d}(k) \slashed{n}\gamma_5(\slashed{p}-\slashed{k})\gamma^\rho (\slashed{p}-\slashed{k}+\slashed{\ell})\gamma_\mu(\slashed{p}_B-\slashed{k}+\slashed{\ell}+m_b)\gamma^{\rho'} b(p_B-k).
\end{align}
The subscript $``b\bar{d}"$ of $\Pi_\mu$ will be omitted from now on. Assigning the external momenta $p$ and $k$ to the  configuration $p^\mu\sim(m_b,\Lambda,0)$ and $k^\mu\sim (\Lambda,\Lambda,\Lambda)$ with $\Lambda\equiv \Lambda_{\text{QCD}}$, we identify the leading-power contribution from the hard-collinear region $\ell^\mu\sim (m_b,\Lambda,\sqrt{m_b\Lambda})$:
\begin{align}
        \Pi_\mu^{(2a)}(p_B,q)
        =&\mu^{4\epsilon}\frac{g_s^4 C_FT_FN_f}{\omega-\bar{n}\cdot p} \frac{(-1)^{-\epsilon}8i}{(4\pi)^{2-\epsilon}}\frac{\Gamma(\epsilon)\Gamma^2(2-\epsilon)}{\Gamma(4-2\epsilon)}  i f_B m_B \phi_{b\bar{d}}^-(\omega) \bar{n}_\mu   \notag\\
        &\times\int \frac{d^D\ell}{(2\pi)^D} \frac{\bar{n}\cdot \ell n\cdot \ell n\cdot p - \ell^2 n\cdot \ell - 2 \ell^2 n\cdot p}{(\ell^2)^{2+\epsilon}n\cdot \ell\left[\ell^2 + n\cdot p \bar{n}\cdot \ell - (\omega-\bar{n}\cdot p )n \cdot \ell - (\omega-\bar{n}\cdot p )n \cdot p\right]  }.
\end{align}

After simplifying the Dirac algebra and evaluating the loop integral in dimensional regularization, we obtain
\begin{align}
        \Pi_\mu^{(2a),hc}(p_B,q)
        =&-C_F  T_F  N_f\left(\frac{\alpha_s}{4\pi}\right)^2\frac{  i f_B m_B \phi_{b\bar{d}}^-(\omega)\bar{n}_\mu  }{\omega-\bar{n}\cdot p}
        \notag
        \\
        &\times \frac{2}{3}\left\{ \frac{1}{\epsilon^3} + \frac{2}{\epsilon^2} \left(L_{\omega}+\frac{4}{3}\right)+\frac{2}{\epsilon}\left(L_{\omega}^2+
        \frac{8}{3} L_{\omega} - \frac{\pi^2}{12} +\frac{44}{9}\right)\right.\notag
        \\
        & \left.\quad+\frac{4}{3} \left[L_{\omega}^3+4L_{\omega}^2-\left(\frac{\pi^2}{4}-\frac{44}{3}\right)L_{\omega} -8\zeta(3)-\frac{\pi^2}{3}+\frac{206}{9} \right] \right\}.\label{eq:2loop2a}
\end{align}

It should be emphasized that the large-$\beta_0$ approximation does not reproduce the complete infrared structure of the full NNLO correction.
In particular, while the present pNNLO result contains poles up to at most $1/\epsilon^3$, the full NNLO result is expected to contain up to $1/\epsilon^4$ terms from additional sources, such as non-factorizing soft-collinear kinematics.
Therefore, the present result should be interpreted only as a partial NNLO contribution in the large-$\beta_0$ limit.
Even with this limitation, it provides a definite and calculable two-loop contribution to the hard-collinear kernel, which isolates the running-coupling-induced effect in the LCSR prediction.

For convenience, we introduce the shorthand notation
\begin{align}
    &\bar{\eta}=1-\eta=1-\frac{\omega}{\bar{n}\cdot p},~~~~
    L_{\bar{\eta}}\equiv\ln{\left(\frac{\bar{n}\cdot p-\omega}{\bar{n}\cdot p}\right)},~~~~
    L_{\eta}\equiv \ln{\left(\eta \right)}\ln^2{(\bar{\eta})}+2{\rm Li}_3(\bar{\eta}),
    \notag\\
    &L_{\omega}(\mu)\equiv\ln{\left[\frac{\mu^2}{n\cdot p\left(\omega-\bar{n}\cdot p\right)}\right]},~~~~
    L_{p}(\mu)\equiv L_{\omega}(\mu)+L_{\bar{\eta}}=\ln{\left(\frac{\mu^2}{-p^2}\right)}.
    \label{eq:notation2}
\end{align}
Here the logarithmic functions are understood for $\bar{n}\cdot p<0$. With this choice, the correlator $\Pi_\mu(p_B,q)$ stays away from the physical branch cut.

The correlation function from QCD correction at $\mathcal{O}(\alpha_s^2\beta_0)$ to the pion vertex is
    \begin{align}
  \Pi_\mu^{(2b)}(p_B,q)=&\mu^{4\epsilon} \frac{g_s^4 C_FT_FN_f}{(p-k)^2}\frac{8i}{(4\pi)^{2-\epsilon}} \frac{\Gamma(\epsilon)\Gamma^2(2-\epsilon)}{\Gamma(4-2\epsilon)}\notag\\
        &\times\int\frac{d^D\ell}{(2\pi)^D} \frac{(\ell^2 g_{\rho\rho'}-\ell_\rho \ell_{\rho'})\bar{d}(k) \gamma^\rho (\slashed{k}-\slashed{\ell}) \slashed{n} \gamma_5 (\slashed{p}-\slashed{k}+\slashed{\ell}) \gamma^{\rho'} (\slashed{p}-\slashed{k}) \gamma_\mu  b(p_B-k)}{[(p-k+\ell)^2+i0][(\ell-k)^2+i0] [\ell^2+i0]^2 (-\ell^2)^\epsilon}.\label{eq:2loop 2b}
    \end{align}
After performing the expansion by regions, we find that the hard region generates a scaleless integral in the dimensional regularization scheme, while the soft and collinear regions are both power suppressed. Therefore, the leading power contribution of $\Pi_{\mu}^{(2b),hc}$ can be derived from Eq.\eqref{eq:2loop 2b} straightforwardly:
{\allowdisplaybreaks
\begin{align}
        &\Pi_\mu^{(2b),hc}(p_B,q)\notag\\
        =& C_F  N_f T_F \left(\frac{\alpha_s}{4\pi}\right)^2\frac{i f_B  m_B \phi^+_{b\bar{d}}(\omega)n_\mu}{\omega-\bar{n}\cdot p}\times\frac{4\bar{\eta}}{3\eta}\left\{\frac{L_{\bar{\eta}}}{\epsilon}+\left[L^2_{\bar{\eta}}+2L_\omega(\mu)L_{\bar{\eta}}+\frac{25}{6}L_{\bar{\eta}}+{\rm Li}_2(\eta)\right]\right\}\notag\\
        +&C_F N_f T_F\left(\frac{\alpha_s}{4\pi}\right)^2\frac{ if_B  m_B\phi_{b\bar{d}}^-(\omega) \bar{n}_\mu }{\omega-\bar{n}\cdot p} \cdot \notag \\
        &\times \frac{4}{3\eta}\left\{ \frac{L_{\bar{\eta}}+\eta }{\epsilon^2}  +\frac{1}{\epsilon}\left[
        L_{p}^2-L_{\omega}^{2}+2\eta L_{\omega}+\left(\frac{17}{3}-\eta\right)L_{\bar{\eta}}+{\rm Li}_2\left(\eta\right)+\frac{59}{12}\eta\right]\right.\notag\\
        &\quad+\frac{2}{3}L_{p}^3+\left(\frac{17}{3}-\eta\right)L_{p}^2-\frac{2}{3}L_{\omega}^3+\left(3\eta-\frac{17}{3}\right)L_{\omega}^{2}+\left(2{\rm Li}_2\left(\eta\right)+\frac{59}{6}\eta\right)L_{\omega}\notag\\
        &\quad+\left(\frac{3\pi^2+428-75\eta}{18}\right)L_{\bar{\eta}}+\left(\frac{17}{3}-\eta\right){\rm Li}_2\left(\eta\right)-{\rm Li}_3\left(\eta\right)-L_\eta\notag\\
        &\quad\left.+\left(\frac{1415}{72}-\frac{\pi^2}{6}\right)\eta+ 2\zeta_3\right\}.\label{eq:2loop 2b2}
\end{align} }
Following the same procedure as before, we can obtain the results for the self-energy diagram Fig.\ref{fig:self} and the box diagram Fig.\ref{fig:box}:
    \begin{align}
        \Pi_\mu^{(2c),hc}(p_B,q)
        =&C_F N_f T_F\left(\frac{\alpha_s}{4\pi}\right)^2\frac{i f_B m_B \phi_{b\bar{d}}^-(\omega)\bar{n}_\mu}{\omega-\bar{n}\cdot p}
        \times \left\{\frac{1}{\epsilon}+2\left(  L_{\omega}+\frac{7}{4}  \right)\right\},\label{eq:2loop2c}
    \end{align}
and
    \begin{align}
        &\Pi_\mu^{(2d),hc}(p_B,q)\notag\\
        =& C_F N_f T_F \left(\frac{\alpha_s}{4\pi}\right)^2\frac{ if_B m_B \phi_{b\bar{d}}^+(\omega)\bar{n}_\mu}{\omega-\bar{n} \cdot p}\cdot \frac{n\cdot p}{m_b}\cdot\frac{4\bar{\eta}}{3\eta}\left\{\frac{L_{\bar{\eta}}}{\epsilon}+\left[L_{p}^2-L_{\omega}^{2}+\frac{25}{6}L_{\bar{\eta}}+{\rm Li}_2\left(\eta\right)\right]\right\}\notag\\
        -&C_F N_f T_F \left(\frac{\alpha_s}{4\pi}\right)^2\frac{ if_B m_B \phi_{b\bar{d}}^-(\omega)\bar{n}_\mu}{\omega-\bar{n} \cdot p}\cdot\frac{4}{3}\left\{\frac{1}{\epsilon^2}\left(\frac{\bar{\eta}L_{\bar{\eta}}}{\eta}+\frac{1}{2}\right)\right.\notag\\
        &+\frac{1}{\epsilon}\left[\frac{\bar{\eta}}{\eta}\left(L_{p}^2-L_{\omega}^{2}+ \frac{11}{3}L_{\bar{\eta}}+{\rm Li}_2\left(\eta\right)\right)+L_{\omega}+\frac{4}{3}\right]\notag\\
        &+\frac{\bar{\eta}}{3\eta}\left(2L_{p}^3+11L_{p}^2-2L_{\omega}^3-11L_{\omega}^{2}+6{\rm Li}_2(\eta)L_{\omega}+\frac{3\pi^2+224}{6}L_{\bar{\eta}}+11{\rm Li}_2(\eta)+6\zeta_3\right)\notag\\
        &\left.-\frac{\bar{\eta}}{\eta}\left(L_\eta+{\rm Li}_3(\eta)\right)+L_{\omega}^{2}+\frac{8}{3}L_{\omega}-\frac{3\pi^2-176}{36}\right\}.\label{eq:2loop2d}
    \end{align}

After computing the two-loop diagrams shown in Fig.\ref{fig:two_loop feynnman diagrams}, we should also include the corresponding contribution
from the counterterm associated with the gluon field renormalization.
This counterterm contribution requires the $\mathcal{O}(\epsilon)$ result of the one-loop correlator $\Pi^{(1)}_{\mu}$. The result reads
\cite{Wang:2015vgv}:
\begin{align}
        &\Pi_{\mu}^{(1),hc}(p_B,q)\notag\\
        =&\frac{\alpha_s C_F}{4\pi}\cdot \frac{if_Bm_B\phi_{b\bar{d}}^-(\omega)\bar{n}_\mu}{\omega-\bar{n}\cdot p} \left\{\frac{2}{\epsilon^2} +\frac{2}{\epsilon}\left(L_{p}-2L_{\bar{\eta}}\right) +\left[ 2L_{\omega}^2-L_{p}^2-\left(\frac{2}{\eta}+1\right)L_{\bar{\eta}}-\frac{\pi^2}{6}-1  \right]\right.\notag\\
        &+\epsilon\left[\frac{2}{3}L_{\omega}^3-\frac{1}{3}L_{p}^3 -\left(\frac{1}{\eta}+\frac{1}{2}\right)\left(L_{p}^2-L_{\omega}^{2}\right) -\left(\frac{\pi^2}{6}+1\right)L_{\omega} \right.\notag\\
        &\left.\left.~~~~-\left(\frac{5}{\eta}-\frac{\pi^2}{6}+3  \right)L_{\bar{\eta}} -\frac{14}{3}\zeta(3)-2\right]\right\}\notag\\
        -&\frac{\alpha_s C_F}{4\pi}\cdot \frac{if_Bm_B\phi_{b\bar{d}}^+(\omega)}{\omega-\bar{n}\cdot p}\cdot \left(n_\mu+\frac{n\cdot p}{m_b} \bar{n}_\mu\right)\frac{\bar{\eta}}{\eta}\left\{ L_{\bar{\eta}}+\epsilon\left[\frac{L_{p}^2}{2}-\frac{L_{\omega}^{2}}{2}+L_{\bar{\eta}}\right] \right\}.
        \label{eq:one loop result}
\end{align}
Since the correlation function $\Pi_{\mu}$ is defined using a conserved current in QCD, no additional operator renormalization is required. Combining the result in Eq.\eqref{eq:one loop result} with the contributions from the four fermion-loop diagrams in Eqs.\eqref{eq:2loop2a}, \eqref{eq:2loop 2b2}, \eqref{eq:2loop2c} and \eqref{eq:2loop2d}, we obtain the renormalized hard-collinear part of the two-loop correlator:
\begin{align}
        \Pi_{\mu}^{(2),hc}&=\Pi_{\mu}^{(2a),hc}+\Pi_{\mu}^{(2b),hc}+\Pi_{\mu}^{(2c),hc}+\Pi_{\mu}^{(2d),hc}+{\frac{\alpha_s}{4\pi}}\frac{4}{3\epsilon}N_f T_F\Pi_{\mu}^{(1),hc}
 \notag\\
        &=
  C^{(0)}\cdot J^{(2)}\otimes \Phi^{(0)}_{b\bar{d}}, \label{eq:2loophctot}
    \end{align}
where the partonic distribution amplitude of the $B$-meson is defined as
\begin{align}
        \Phi_{b \bar{d}}^{(0) }\left(\omega^{\prime}\right)=\delta\left(\bar{n} \cdot k-\omega^{\prime}\right) \bar{d}_{\beta}(k) b_{\alpha}\left(p_{B}-k\right).
\end{align}
From the finite part of the above expression, one can extract the two-loop jet function. Substituting $N_f \to -\frac{3}{2}\beta_0$, $\phi^{\pm}_{b\bar{d}}(\omega)\to \phi^{\pm}_B(\omega)$ and inserting the resulting jet function into Eq.\eqref{eq:9},

we arrive at the RG-improved LCSR for the $B \to \pi$ form factors at leading power, including the $\mathcal{O}(\alpha_s^2\beta_0)$ contribution:
\begin{align}
&f_{\pi} \exp \left[-\frac{m_{\pi}^{2}}{n \cdot p \, \omega_{M}}\right]\left\{\frac{n \cdot p}{m_{B}} f_{B\pi}^{+}\left(q^{2}\right), f_{B\pi}^{0}\left(q^{2}\right)\right\} \notag
\\
&\quad=\mathcal{F}_B(\mu) \int_{0}^{\omega_{s}} d \omega^{\prime} e^{-\omega^{\prime} / \omega_{M}}
\left\{\left[\mathcal{C}_{\bar{n}}^{(+)}\left(n \cdot p, \mu_{h 1}\right)\boldsymbol{\Phi}_{\bar{n}}^{+}\left(\omega^{\prime}, \mu\right)
+ \mathcal{C}_{\bar{n}}^{(-)}\left(n \cdot p, \mu_{h 1}\right) \boldsymbol{\Phi}_{\bar{n}}^{-}\left(\omega^{\prime}, \mu\right) \right] \right. \notag
\\
&\left.\quad \, \pm \frac{n \cdot p-m_{B}}{m_{B}}\left[\mathcal{C}_{n}^{(+)}\left(n \cdot p, \mu_{h 1}\right)\boldsymbol{\Phi}_{n}^{+}\left(\omega^{\prime}, \mu\right)
+ \mathcal{C}_{n}^{(-)}\left(n \cdot p, \mu_{h 1}\right) \boldsymbol{\Phi}_{n}^{-}\left(\omega^{\prime}, \mu\right) \right]\right\},\label{eq:NNLLsumrule}
\end{align}
with renormalization group improved $\mathcal{C}(n\cdot p,\mu)$ and $\mathcal{F}_B(\mu)$ defined as,
\begin{align}
    &\mathcal{C}^+_{n,\bar{n}}(n\cdot p,\mu)\equiv U_{1}\left(n \cdot p, \mu_{h 1}, \mu\right) C_{n,\bar{n}}^{(+)}(n\cdot p,\mu_{h1}), \quad \mathcal{F}_B(\mu)\equiv U_{2}\left(\mu_{h 2}, \mu\right) \tilde{f}_{B}\left(\mu_{h 2}\right), \notag \\
    &\mathcal{C}^-_{n,\bar{n}}(n\cdot p,\mu)\equiv U_{3}\left(n \cdot p, \mu_{h 1}, \mu\right) C_{n,\bar{n}}^{(-)}(n\cdot p,\mu_{h1}).
\end{align}
The effective distribution amplitudes $\boldsymbol{\Phi}_{\bar{n},n}^{\pm}(\omega')$, which incorporate the hard-collinear dynamics, are obtained by taking the imaginary part of the jet function. The NLO expressions for $J_{n,\bar{n}}^{\pm}$ and $\boldsymbol{\Phi}_{\bar{n},n}^{\pm}(\omega')$, which enter the NLL-improved sum rules for $f_{B\pi}^{+,0}$, have been collected in Refs.~\cite{Wang:2015vgv,Cui:2022zwm}.
The additional ingredients of $J_{n,\bar{n}}^{\pm}$ and $\boldsymbol{\Phi}_{\bar{n},n}^{\pm}(\omega')$ required for the $\mathcal{O}(\alpha_s^2\beta_0)$ contribution are presented in Appendix~\ref{app:2loopresults}.

\section{Numerical analysis}
After incorporating the available leading-power contributions up to $\mathcal{O}(\alpha_s^2\beta_0)$ and the tree-level next-to-leading-power corrections, we obtain the LCSR predictions for the $B\to\pi$ form factors in the large-recoil region. {For the latter,
we employ the expressions derived in Ref.~\cite{Cui:2022zwm}, including the contributions from the power expansion of the hard-collinear propagator, the subleading heavy-light current, and the higher-twist $B$-meson LCDAs.}
We then study the numerical impact of this partial NNLO correction on the form factors. Combining our results with the LQCD inputs at low recoil, we perform a fit to the coefficients of the BCL parametrization~\cite{Bourrely:2005hp,Bourrely:2008za,Lellouch:1995yv}, which allows us to extend the form factors to the full kinematic region.

As a brief phenomenological application, we also study the impact of the present partial NNLO result on the extraction of the CKM matrix element $|V_{ub}|$.
We further analyze phenomenological observables, including the decay branching ratios, the lepton-flavour-universality ratio $R_\pi$, the forward-backward asymmetry $A^{B\to\pi \ell\bar{\nu}_\ell}_{\text{FB}}$, the polarization asymmetry $A^{B\to\pi \ell\bar{\nu}_\ell}_{\lambda_\ell}$, and the flat term $F^{B\to\pi \ell\bar{\nu}_\ell}_{\text{H}}$.
Since the partial NNLO correction only leads to a modest quantitative update and does not qualitatively change the phenomenological picture obtained in previous analyses~\cite{Cui:2022zwm}, we keep the discussion concise.

    \subsection{Input parameters}

    \begin{table}[ht]
        \footnotesize
        \centering \setlength\tabcolsep{6pt} \def\arraystretch{1.5}
        \caption{Numerical values of the input parameters.}
        \vspace{3pt}
        \begin{tabular}{|l|ll||l|ll|}
            \hline\hline
            Parameters & Values & Ref. & Parameters & Values & Ref.\\
            \hline\hline
            $m_{B^\pm}$ & 5.27942(8)~GeV & \cite{ParticleDataGroup:2024cfk} & $m_{B^*}$ & 5.32475(20)~GeV & \cite{ParticleDataGroup:2024cfk}\\
            $m_{B^0}$ & 5.27963(20)~GeV & \cite{ParticleDataGroup:2024cfk} & $m_{\pi^0}$ & 134.9768(5)~MeV & \cite{ParticleDataGroup:2024cfk}\\
            $m_{\pi^\pm}$ & 139.57039(18)~MeV & \cite{ParticleDataGroup:2024cfk} & $m_{\mu}$ & 105.658~MeV & \cite{ParticleDataGroup:2024cfk}\\
            $m_b(m_b)$ & $4.200(14)$~GeV &  \cite{Beneke:2014pta} & $m_{\tau}$ & 1776.93(9)~MeV & \cite{ParticleDataGroup:2024cfk}\\
            $G_F$ & $1.166379\times10^{-5}$~GeV$^{-2}$ &  \cite{ParticleDataGroup:2024cfk} & $\alpha_s^{(5)}(m_Z)$ & 0.1180$\pm$0.0009 & \cite{ParticleDataGroup:2024cfk}\\
            \hline\hline
            $f_{B^0}$ & 190.0(1.3)~MeV & \cite{FlavourLatticeAveragingGroupFLAG:2021npn} & $f_{\pi^\pm}$ & 130.2(0.8)~MeV & \cite{FlavourLatticeAveragingGroupFLAG:2021npn}\\
            \hline\hline
            $\mu_{ h1}$ & $\left[m_b/2,2m_b\right]$ &  & $\mu_{ h2}$ & $\left[m_b/2,2m_b\right]$ & \\
            $\mu=\mu_{ hc}$ & 1.5(5) ~GeV &  & $\mu_0$ & 1.0~GeV & \\
            \hline\hline
            $M^2$ & $1.25(25)~{\rm GeV^2}$ & \cite{Wang:2015vgv} & $s_0$ & $0.70(5)~{\rm GeV^2}$ & \cite{Wang:2015vgv}\\
            \hline\hline
            $\lambda_B(\mu_0)$ & 0.35(15)~GeV & \cite{Cui:2022zwm} &  & $\left\{0.7,6.0\right\}$ & \\
            $\lambda_E^2(\mu_0)/\lambda_H^2(\mu_0)$ & 0.5(1) & \cite{Cui:2022zwm} & $\left\{ \hat{\sigma}_1(\mu_0),\hat{\sigma}_2(\mu_0) \right\}$ & $\left\{0.0,\pi^2/6\right\}$ & \cite{Cui:2022zwm}\\
            $2\lambda_E^2(\mu_0)+\lambda_H^2(\mu_0)$ & $0.25(15)~{\rm GeV^2}$ & \cite{Cui:2022zwm} &  & $\left\{-0.7,-6.0\right\}$ & \\
            \hline\hline
        \end{tabular}
        \label{tab:input parameters}
    \end{table}

 In Table~\ref{tab:input parameters}, we summarize the numerical values of the input parameters, with uncertainties given in parentheses.
We adopt the numerical results of the five-loop evolution of the QCD coupling constant $\alpha_s(\mu)$ and the $b$-quark mass $m_b(\mu)$ in the $\overline{\rm MS}$ scheme by using the \textbf{RunDec} package \cite{Chetyrkin:2000yt}.
Two hard-matching scales $\mu_{h1}$ and $\mu_{h2}$  introduced in the hard functions and the $B$-meson decay constant, respectively, are varied within the interval $[m_b/2, 2m_b]$.
We set the factorization scale to the same range as the hard-collinear scale $\mu=\mu_{\rm hc}=1.5\pm0.5~ {\rm GeV}$.

\begin{figure}[t]
    \centering
    \includegraphics[width=0.9\textwidth]{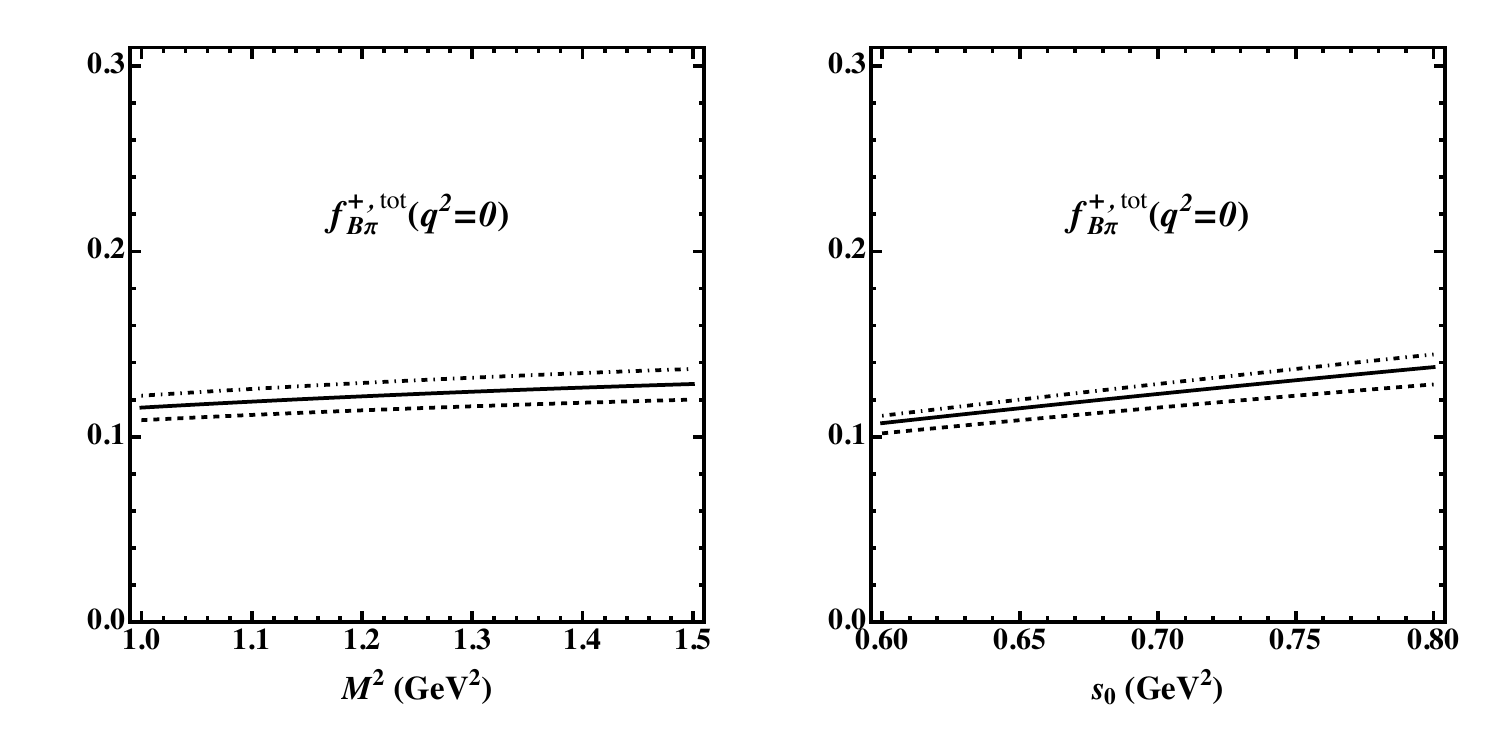}
    \vspace{-10pt}
    \caption{
    Dependence of the form factor $f^+_{B\pi}(0)$ on the Borel parameter
    $M^2$ and on the effective threshold $s_0^\pi$. The result is computed
    from the LCSR including the available leading-power NLO correction, the
    known subleading-power contributions, and the partial NNLO correction of
    order $\mathcal{O}(\alpha_s^2\beta_0)$. In the left panel, the solid,
    dashed and dot-dashed curves correspond to
    $s_0^\pi=0.70\,{\rm GeV}^2$, $0.65\,{\rm GeV}^2$ and
    $0.75\,{\rm GeV}^2$, respectively. In the right panel, the solid,
    dashed and dot-dashed curves correspond to
    $M^2=1.25\,{\rm GeV}^2$, $1.00\,{\rm GeV}^2$ and
    $1.50\,{\rm GeV}^2$, respectively.
    }
    \label{fig:borel-threshold-stability}
\end{figure}

The Borel parameter $M^2$ and the effective threshold $s_0^\pi$ are auxiliary
quantities introduced by the Borel transformation and the continuum subtraction
in the pion channel. In the notation of the dispersion relation in the variable
$\bar n\cdot p$, they are related to the corresponding SCET sum-rule parameters
by $M^2=(n\cdot p)\,\omega_M$ and $s_0^\pi=(n\cdot p)\,\omega_s$.
Since these parameters are auxiliary quantities of the sum-rule construction,
their variation provides an important internal check of the numerical analysis.
In Fig.\ref{fig:borel-threshold-stability}, we show the dependence of $f^+_{B\pi}(0)$ on the Borel parameter and on the effective threshold, after including the known subleading-power contributions and the partial NNLO correction.
We find that the inclusion of the $\mathcal{O}(\alpha_s^2\beta_0)$ term does not lead to a significant shift of the preferred stability region used in Ref.~\cite{Wang:2015vgv}.
The same working windows,
\begin{equation}
    M^2\in[1.0,1.5]\,{\rm GeV}^2,\qquad
    s_0^\pi\in[0.65,0.75]\,{\rm GeV}^2 ,
\end{equation}
remain appropriate for the numerical analysis. The residual dependence on these auxiliary parameters is included in the uncertainty estimate.
We stress, however, that this observation should not be interpreted as a proof of perturbative stability of the full sum rule, because the present $\mathcal{O}(\alpha_s^2\beta_0)$ contribution represents only a partial NNLO correction and does not include the complete two-loop hard-collinear kernel or radiative corrections involving three-particle $B$-meson LCDAs.

{We adopt the three-parameter model for the universal $B$-meson LCDAs introduced in Ref.~\cite{Beneke:2018wjp}},
which satisfies the constraints from the equations of motion and has the expected behavior in the small-momentum region.{
In this parametrization, the leading-twist
two-particle LCDA $\phi_B^+(\omega,\mu_0)$ is given by
\begin{equation}
    \phi_B^+(\omega,\mu_0)=
U( \beta-\alpha,3-\alpha,\omega/\omega_0 ) \,
\frac{\omega}{\omega_0^2}
\frac{\Gamma(\beta)}{\Gamma(\alpha)}\,
e^{-\frac{\omega}{\omega_0}},
\end{equation}
while the two-particle twist-three LCDA $\phi_B^-(\omega,\mu_0)$ is decomposed as
\begin{equation}
    \phi_B^-(\omega,\mu_0)
    =
    \phi_B^{-\rm WW}(\omega,\mu_0)
    +
    \phi_B^{-\rm t3}(\omega,\mu_0).
\end{equation}
Here the Wandzura-Wilczek contribution $\phi_B^{-\rm WW}$ is determined by the leading-twist
LCDA $\phi_B^+$ through the equation-of-motion relation
\begin{equation}
    \phi_B^{-\rm WW}(\omega,\mu_0)
    =
    \int_\omega^\infty \frac{d\eta}{\eta}\,
    \phi_B^+(\eta,\mu_0),
\end{equation}
whereas $\phi_B^{-\rm t3}$ denotes the genuine twist-three contribution induced by the three-particle LCDA $\Phi_3$ and is proportional to $\lambda_E^2-\lambda_H^2$ in the adopted model.
Within this parametrization, the available leading-logarithmic renormalization-group evolution of $\phi_B^{\pm}(\omega,\mu)$ can be implemented in analytic form, as summarized in Appendix~\ref{app-B-LCDA}. We stress, however, that this treatment does not constitute the complete coupled RG evolution of the full two- and three-particle higher-twist LCDA system. In particular, the explicit mixing of $\phi_B^-$ with three-particle LCDAs is not included in our perturbative analysis.
}
The details are provided in Appendix \ref{app-B-LCDA}. The parameters $\alpha$, $\beta$ and $\omega_0$ in this model are related to the inverse moments $\lambda_B$ and $\hat{\sigma}_{1,2}$ of the leading-twist $B$-meson LCDA as follows,
\begin{equation}
\begin{aligned}
&\lambda_B(\mu)=\frac{\alpha-1}{\beta-1}\omega_0,
\\
&\hat{\sigma}_{1}(\mu)=\psi(\beta-1)-\psi(\alpha-1)+\ln{\frac{\alpha-1}{\beta-1}},\\
&\hat{\sigma}_{2}(\mu)=\hat{\sigma}_{1}^2(\mu)-\psi'(\beta-1)+\psi'(\alpha-1)+\frac{\pi^2}{6},
\end{aligned}
\end{equation}
where $\psi(x)$ is the digamma function and
\begin{equation}
\begin{aligned}
 &\frac{1}{\lambda_B(\mu)}=\int_0^\infty \frac{d w }{w} \phi_B^+(w,\mu),
 \\
&\frac{\hat{\sigma}_{n}(\mu)}{\lambda_B(\mu)}=\int_0^\infty \frac{d w }{w} \ln^n\frac{e^{-\gamma_E}\lambda_B(\mu)}{w} \phi_B^+(w,\mu).
\end{aligned}
\end{equation}
The one-loop evolution of these moments reads
\begin{align}
\frac{\lambda_B(\mu)}{\lambda_B(\mu_0)}
&= 1+ \frac{\alpha_s(\mu_0)\,C_F}{\pi}\ln\frac{\mu}{\mu_0}
\left[\hat{\sigma}_1(\mu_0)+\ln\frac{\sqrt{\mu\mu_0}e^{\gamma_E}}{\lambda_B(\mu_0)}-\frac{1}{2} \right] ,
\nonumber \\
\hat{\sigma}_1(\mu)
&= \hat{\sigma}_1(\mu_0)
+ \frac{\alpha_s(\mu_0)\,C_F}{4\pi}\,4\ln\frac{\mu}{\mu_0}
\left[\hat{\sigma}^2_1(\mu_0)-\hat{\sigma}_2(\mu_0) \right] .
\end{align}
We neglect the scale dependence of $\hat{\sigma}_2(\mu)$ due to the expected small effect.{
Substantial progress has been made in the theoretical determination of both light- and heavy-meson LCDAs
\cite{LatticeParton:2022zqc,LatticeParton:2024zko,Han:2024min,Khodjamirian:2020hob,Janowski:2021yvz,Mandal:2023lhp}.
Nevertheless, a precise first-principles determination of the $B$-meson inverse moment $\lambda_B$ remains challenging.}
We prefer the conservative interval $\lambda_B(\mu_0)=0.35(15)$ GeV with large uncertainty \cite{Wang:2018wfj,Janowski:2021yvz,Cui:2022zwm}. We take $\left\{ \hat{\sigma}_1(\mu_0),\hat{\sigma}_2(\mu_0) \right\}=\{0.0,\pi^2/6\}$ as the default choice for LCDAs, and the model dependent uncertainties of the form factors are included by varying the values of $\left\{ \hat{\sigma}_1(\mu_0),\hat{\sigma}_2(\mu_0) \right\}$ in the final results.
    The HQET parameters $\lambda_E^2$ and $\lambda^2_H$ appear in the three-body $B-$meson matrix element,
\begin{equation}
\begin{aligned}
 \langle 0| \bar q(0) g_s G_{\mu\nu}(0)\Gamma h_v(0)|\bar B(v)\rangle =&-\frac{i}{6} \tilde{f}_B(\mu)m_B \lambda^2_H {\rm{Tr}}\left[\gamma_5\Gamma P_+ \sigma_{\mu\nu}\right]
\\
&-\frac{1}{6} \tilde{f}_B(\mu)m_B\left( \lambda^2_H- \lambda^2_E\right)
  {\rm{Tr}}\left[\gamma_5\Gamma P_+(v_\mu\gamma_\nu-v_\nu\gamma_\mu)\right],
\end{aligned}
\end{equation}
whose values can be determined through QCD sum rules \cite{Grozin:1996pq, Nishikawa:2011qk,Rahimi:2020zzo}.

Due to the discrepancies among different theoretical determinations, which may be attributed to different treatments of higher-order and higher-power corrections, we take the intervals for $2\lambda_E^2+\lambda^2_H$ and $\lambda_E^2/\lambda_H^2$ displayed in Table~\ref{tab:input parameters}.
These intervals cover the ranges of the two theoretical calculations~\cite{Grozin:1996pq,Nishikawa:2011qk} and simultaneously satisfy the upper bounds in Ref.~\cite{Rahimi:2020zzo}.

\subsection{Numerical results for the \texorpdfstring{$B \to \pi $}{B to pi} form factors}

In Table~\ref{tab:FF at q=0}, we decompose the LCSR prediction for the $B\to\pi$ form factors at $q^2=0$ into the LO, NLO, NLP and partial NNLO contributions.
Both the NLO and NLP corrections give sizable negative shifts to the form factors $f_{B\pi}^{+,0}(q^2)$, at the level of about $-30\%$ relative to the LO contribution.
The pNNLO correction gives a positive contribution, approximately $+6.1\%$ relative to the LO result.
Numerically, this pNNLO effect is much smaller than the NLP contribution and should not be interpreted as the dominant missing correction in the present prediction.

\begin{table}[ht]
\centering \setlength\tabcolsep{6.0pt} \def\arraystretch{1.5}
\caption{Decomposition of the LCSR prediction for $f^{+,0}_{B\pi}(0)$ into the LO, NLO, NLP and pNNLO contributions. The pNNLO term denotes the $\mathcal{O}(\alpha_s^2\beta_0)$ contribution.}
\vspace{3pt}
\begin{tabular}{|c||c|c|c|c|c|}
\hline\hline
& Total  & LO & NLO & NLP  & pNNLO  \\  \hline\hline
$f^{+,0}_{B\pi}(0)$
& $0.146^{+0.194}_{-0.099}$
& $0.325^{+0.185}_{-0.102}$
& $-0.099^{+0.020}_{-0.022}$
& $-0.100^{+0.013}_{-0.013}$
& $0.020^{+0.021}_{-0.009}$ \\  \hline\hline
\end{tabular}
\label{tab:FF at q=0}
\end{table}

In Table~\ref{tab:FF-error-breakdown}, we further decompose the uncertainty of the total result and of the individual LO, NLO, NLP and pNNLO contributions according to the input parameters.
The dominant uncertainty comes from the inverse moment $\lambda_B$, which controls the small-$\omega$ behaviour of $\phi_B^-(\omega)$.
This uncertainty is mainly inherited from the LO result, where the convolution is directly sensitive to $\lambda_B$.

\begin{table}[ht]
\centering
\setlength\tabcolsep{4.0pt}
\def\arraystretch{1.5}
\caption{The summary of the sizeable uncertainties for the total result and for the LO, NLO, NLP and pNNLO contributions. Each column gives the uncertainty induced by varying different input parameters.}
\vspace{3pt}
\begin{tabular}{|c||c||c|c|c|c|c|c|c|}
\hline\hline
$ f^{+,0}_{B\pi}(0)$
& Central value
& $\lambda_B$
& $\hat{\sigma}_{1,2}$
& $2\lambda_{E}^{2}+\lambda_{H}^{2}$
& $\lambda_{E}^2/\lambda_{H}^{2}$
& $M^2$
& $s_0^\pi$
& scale  \\
\hline\hline
Total
& $0.146$
& $^{+0.190}_{-0.078}$
& $^{+0.039}_{-0.060}$
& $^{+0.008}_{-0.008}$
& $^{+0.004}_{-0.003}$
& $^{+0.006}_{-0.009}$
& $^{+0.008}_{-0.009}$
& $^{+0.000}_{-0.013}$ \\
LO
& $0.325$
& $^{+0.180}_{-0.082}$
& $^{+0.038}_{-0.054}$
& $^{+0.006}_{-0.006}$
& $^{+0.004}_{-0.003}$
& $^{+0.012}_{-0.017}$
& $^{+0.014}_{-0.015}$
& $^{+0.002}_{-0.013}$ \\
NLO
& $-0.099$
& $^{+0.009}_{-0.006}$
& $^{+0.009}_{-0.003}$
& $^{+0.000}_{-0.000}$
& $^{+0.000}_{-0.000}$
& $^{+0.003}_{-0.002}$
& $^{+0.002}_{-0.002}$
& $^{+0.012}_{-0.021}$ \\
NLP
& $-0.100$
& $^{+0.004}_{-0.005}$
& $^{+0.000}_{-0.011}$
& $^{+0.001}_{-0.001}$
& $^{+0.001}_{-0.001}$
& $^{+0.006}_{-0.004}$
& $^{+0.004}_{-0.004}$
& $^{+0.010}_{-0.002}$ \\
pNNLO
& $0.020$
& $^{+0.021}_{-0.008}$
& $^{+0.004}_{-0.004}$
& $^{+0.001}_{-0.001}$
& $^{+0.000}_{-0.000}$
& $^{+0.001}_{-0.001}$
& $^{+0.001}_{-0.001}$
& $^{+0.000}_{-0.003}$ \\ \hline\hline
\end{tabular}
\label{tab:FF-error-breakdown}
\end{table}

It is useful to distinguish this dominant LO sensitivity from the uncertainties of the higher-order and power-suppressed terms.
The NLO correction is less directly controlled by $\lambda_B$ than the LO contribution, because the loop correction introduces logarithmic kernels and therefore probes logarithmic inverse moments of the LCDA rather than only the simple inverse moment.
The NLP contribution also has a weaker dependence on $\lambda_B$, since the corresponding power-suppressed kernels contain additional factors of $\omega/m_b$, which soften the endpoint sensitivity.
The pNNLO contribution is numerically small compared with the NLP correction and is largely hidden by the much larger LCDA uncertainty.
Consequently, the impact of the pNNLO correction on the subsequent phenomenological observables is modest.

The uncertainty associated with the effective threshold $s_0^\pi$ can be reduced by using a sum-rule ratio in which the pion decay constant $f_\pi$ in the denominator of the $B\to\pi$ LCSR is replaced by its two-point QCD sum-rule expression.
In such a construction, part of the dependence on the pion-channel threshold cancels between the numerator and the denominator.
We have not adopted this procedure in the present work, because the numerical uncertainty induced by $s_0^\pi$ is not the dominant one in our final result and the use of the physical value of $f_\pi$ is also natural.

Reducing the dominant $\lambda_B$ uncertainty is more challenging. At present, a first-principles determination of the $B$-meson LCDA parameters, in particular $\lambda_B$ and the small-$\omega$ behaviour of $\phi_B^-(\omega)$, is still subject to significant systematic uncertainties.
A realistic way forward is to perform a global analysis of several $B$-meson decay channels that depend on the same LCDA parameters, fitting the LCDA parameters and the CKM parameters simultaneously~\cite{Li:2025mhq}.
In this respect, the strong sensitivity of the LO result to $\lambda_B$ is not only a limitation, but may also provide useful constraints on the $B$-meson LCDA in future global analyses.

\begin{figure}[ht]
        \centering
        \includegraphics[width=0.8\textwidth]{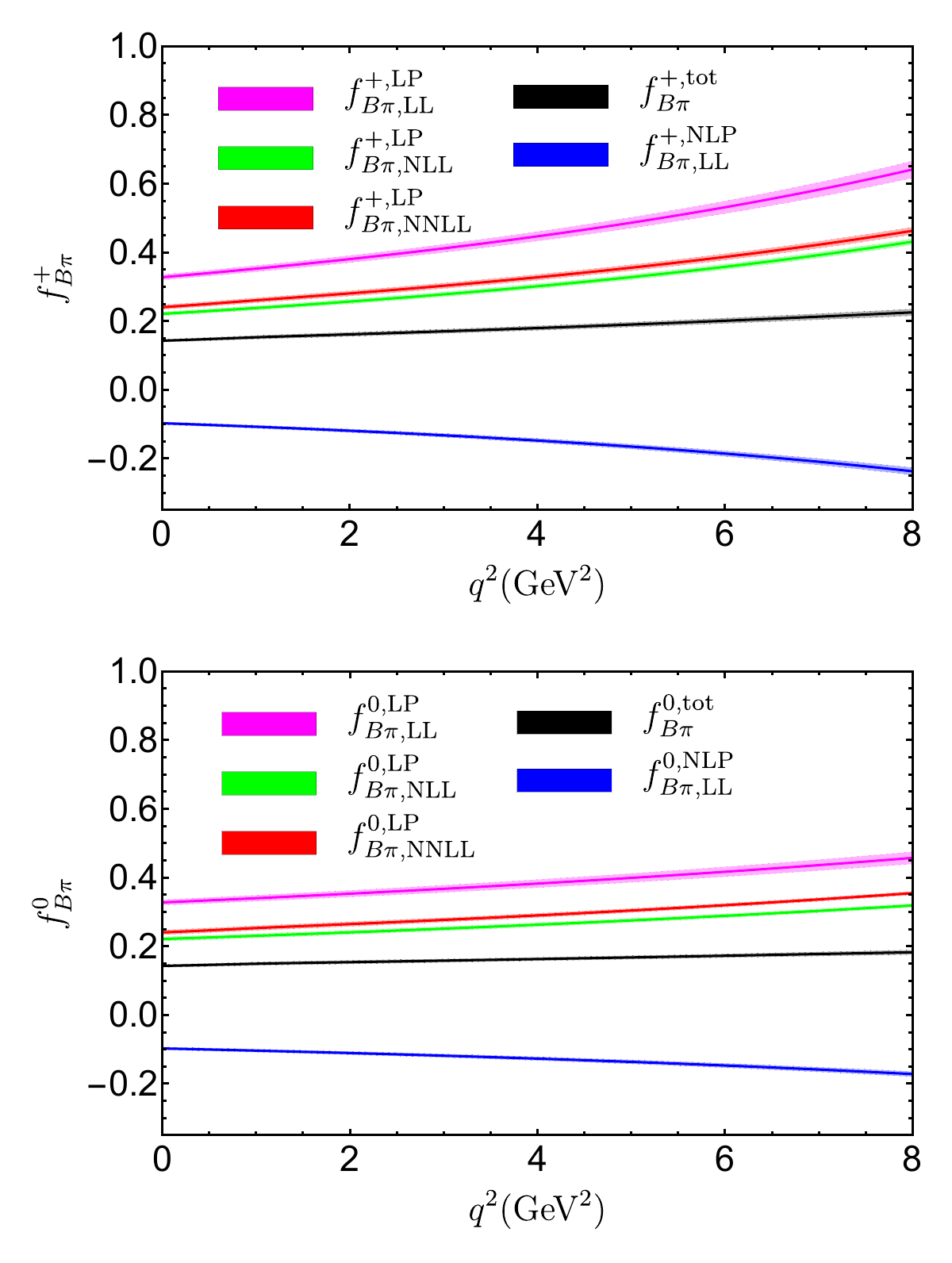}
        \captionsetup{labelfont=bf}
        \vspace{-20pt}
        \caption{Comparison of the leading logarithm resummation improved tree-level contribution, next-to-leading logarithm resummation improved one-loop correction, partial NNLO correction of order $\mathcal{O}(\alpha_s^2\beta_0)$, next-to-leading power correction at tree level and total result to the vector form factor $f_{B\pi}^+$ (upper panel) and scalar form factor $f_{B\pi}^0$ (lower panel) in the kinematic region of $0\leq q^2\leq 8~\rm{GeV}^2$. The shaded bands represent the uncertainties from the variation of hard scale $\mu_{h}$ and factorization scale $\mu$.}
        \label{fig:ffscale}
\end{figure}

In Fig.\ref{fig:ffscale}, we display the $q^2$ dependence of the perturbative and power corrections to the $B\to\pi$ form factors in the large-recoil region $0<q^2<8~{\rm GeV}^2$.
The shaded bands represent the uncertainties estimated from the variation of the perturbative scales.
After including the $\mathcal{O}(\alpha_s^2\beta_0)$ contribution, the residual scale dependence is mildly reduced.

The soft-collinear factorization formula for the vacuum-to-$B$-meson correlation function defined in Eq.(\ref{eq:define the correlation function}) is valid only in the small momentum transfer region $0\lesssim q^2\lesssim 8 \, {\rm GeV}^2$.
In the large momentum transfer region, the LQCD simulations give precise results for the form factors \cite{Colquhoun:2015mfa,Colquhoun:2022atw,Dalgic:2006dt,FermilabLattice:2015cdh,FermilabLattice:2015mwy,Flynn:2015mha}.
In order to obtain the form factors across the entire kinematic region, we perform the combined fit of the LQCD results and LCSR calculations for the form factors using BCL parametrization based upon the positivity and analyticity properties \cite{Bourrely:2008za, Leljak:2021vte, Cui:2022zwm, Han:2023pgf,Gonzalez-Solis:2021pyh}.
We adopt the conformal transformation \cite{Lellouch:1995yv,Bourrely:2005hp,Bourrely:2008za}
\begin{align}
        z(q^2,t_0)=\frac{\sqrt{t_+-q^2}-\sqrt{t_+-t_0}}{\sqrt{t_+-q^2}+\sqrt{t_+-t_0}},
        \label{eq:3.1}
\end{align}
with $t_+\equiv \left(m_B+m_\pi\right)^2$ for the $B\to \pi$ transition form factors. The parameter $t_0$ is usually chosen to be
\begin{align}
        t_0\equiv t_+-\sqrt{t_+(t_+-t_-)},~~~~t_-\equiv (m_B-m_\pi)^2.
\end{align}
This maps the cut region in $q^2$-plane onto the unit circle $|z|=1$ in the $z$-plane, while the physical region is mapped to the real axis segment $-z_{\text{max}}<z<z_{\text{max}}$ with $z_{\text{max}}=0.28$ for the domain $q^2\in\left[0,(m_B-m_\pi)^2\right]$.
Taking into account the asymptotic behaviour of the form factors near the threshold and the scaling behaviour $f_{B\pi}^+(q^2)\sim 1/q^2$ from perturbative QCD, we adopt the simplified $z$-series expansion \cite{Bourrely:2008za},
\begin{align}
        &f_{B\pi}^+(q^2)=\frac{1}{1-q^2/m_{B^*}^2}\sum^{N-1}_{k=0}b_k^+\left[z(q^2,t_0)^k-(-1)^{k-N}\frac{k}{N}z(q^2,t_0)^N\right],\\
        &f_{B\pi}^0(q^2)=\sum^{N-1}_{k=0}b_k^0~z(q^2,t_0)^k.
\end{align}
It is noted that $B^*$ is the only bottom-resonance in the $J^P=1^{-}$ channel located below the branch point. Using the relationship $f_{B\pi}^{+}(0)=f_{B\pi}^0(0)$, we can obtain
\begin{align}
        b_2^0=12.78\left(b_0^+-b_0^0\right)+3.482b_1^++1.186b_2^+-3.575b_1^0,
\end{align}
with the $z$-series expansion truncated at $N=3$.

\begin{figure}[ht]
        \centering
            \includegraphics[width=0.8\textwidth]{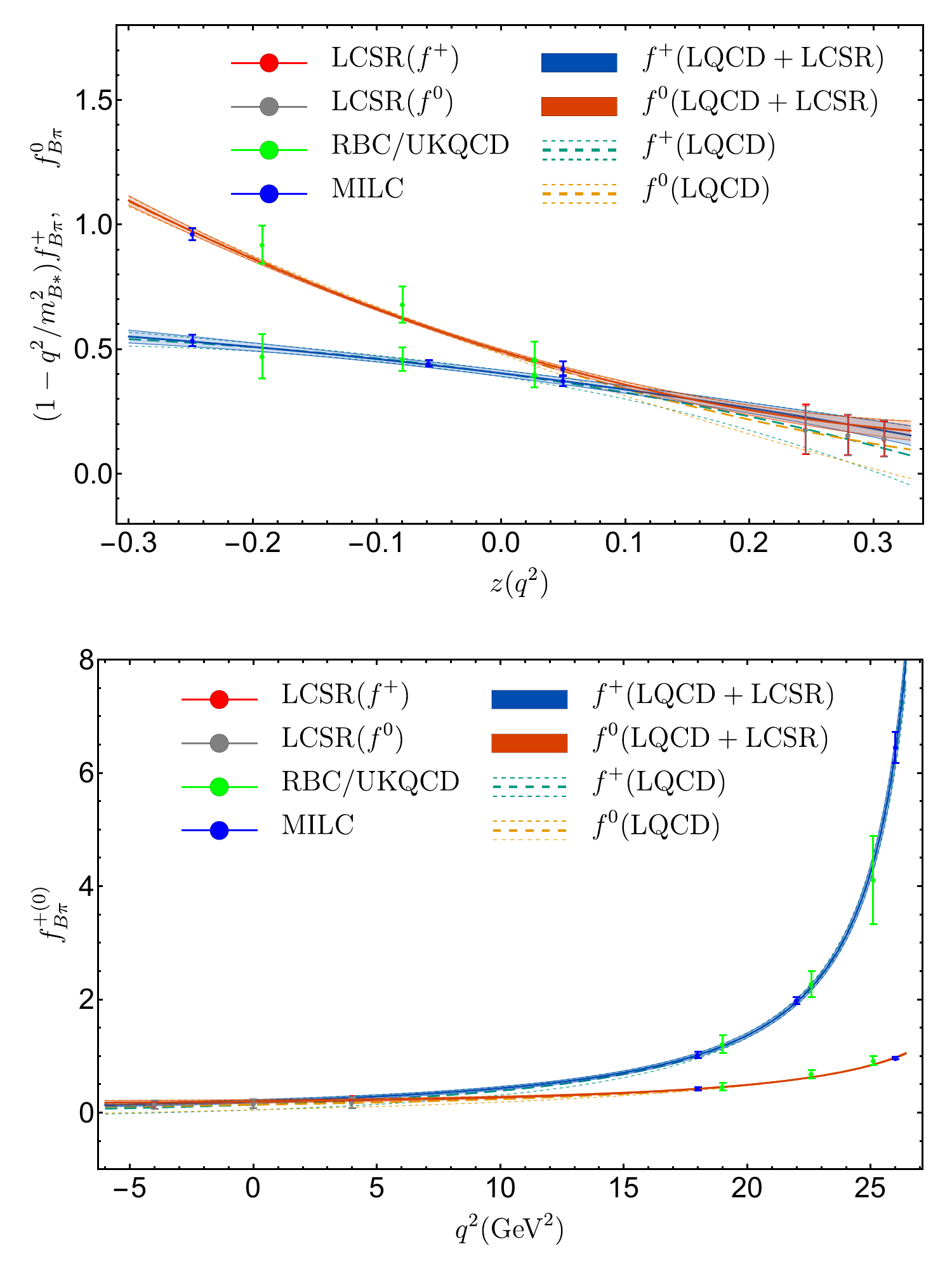}
        \vspace{-20pt}
        \captionsetup{labelfont=bf}
        \caption{The theory prediction for the vector and scalar form factors of $B\to\pi$ transition versus $z$ (upper panel) and versus $q^2$  (lower panel) obtained from the combined fit of the updated LCSR (from this work) and the LQCD (from \cite{Flynn:2015mha,FermilabLattice:2015mwy}) data points in the entire kinematic region. An alternative $z$-series fit for the form factors exclusively using the ``only LQCD'' data points is presented for a comparison.}
        \label{fig:ffallspace}
\end{figure}

We proceed to determine the coefficients of the BCL series $b_k^{+,0}$ through the minimal-$\chi^2$ fit. We will use the updated LCSR predictions for the $B\to\pi$ form factors at three distinct momentum transfers, namely $q^2=\left\{-4,0,4\right\}{\rm GeV^2}$, and incorporate LQCD predictions for the form factors at large $q^2$ region into the fit.
For the LQCD, we use values of the form factors provided by the RBC/UKQCD collaboration at three specific points $q^2=\left\{19.0,~22.6,~25.1\right\}{\rm GeV^2}$ \cite{Flynn:2015mha}, and the values provided by FNAL/MILC collaboration \cite{FermilabLattice:2015mwy} at three points $q^2=\left\{18.0,~22.0,~26.0\right\}{\rm GeV^2}$.
Utilizing the $\chi^2$ fitting approach with 11 degrees of freedom, we have determined the minimum $\chi^2$ value to be $\chi^2/{\rm d.o.f} = 0.59$. It is observed that incorporating the LCSR calculations reduces the errors on the form factors in the large recoil region by $68\%$, as shown in Fig.\ref{fig:ffallspace} . The central values, errors, and correlation matrix for the BCL parameters are presented in Table~\ref{tab:BCL parameters}.

\begin{table}[ht]
        \footnotesize
        \centering\setlength\tabcolsep{6pt} \def\arraystretch{1.5}
            \caption{The central values, errors, and correlation matrix of the $z$-series coefficients for the vector and scalar form factors given by the combined $\chi^2$ fit of LCSR and LQCD results \cite{Flynn:2015mha,FermilabLattice:2015mwy} in the case of $N=3$.}
            \vspace{3pt}
        \begin{tabular}{|c|c||ccccc|}
            \hline\hline
            \multicolumn{2}{|c||}{$B\to \pi$ Form Factors}  & \multicolumn{5}{|c|}{Correlation Matrix} \\
            \hline\hline
            Parameters & Values & $b_0^+$ & $b_1^+$ & $b_2^+$ & $b_0^0$ & $b_1^0$ \\
            \hline\hline
            $b_0^+$ & 0.403(12) & 1 & 0.284 & -0.410 & 0.051 & 0.299 \\
            $b_1^+$ & -0.614(69) &  & 1 & -0.138 & 0.144 & 0.622 \\
            $b_2^+$ & -0.418(257) &  &  & 1 & 0.221 & 0.415\\
            $b_0^0$ & 0.492(8) &  &  &  & 1 & 0.484 \\
            $b_1^0$ & -1.517(61) &  &  &  &  & 1 \\
            \hline\hline
        \end{tabular}
        \label{tab:BCL parameters}
\end{table}

\subsection{Phenomenological analysis of the \texorpdfstring{$B \to \pi \ell \bar{\nu}_{\ell}$}{B to pi ell nu} observables}

We investigate several phenomenological observables in the semileptonic $B \to \pi \ell \bar{\nu}_{\ell}$ decays.
These observables are constructed from the full angular distribution, including the differential branching fraction, the normalized forward-backward asymmetry, the ``flat term'', the lepton-flavour-universality ratio, and the lepton polarization asymmetry.
We first give the full decay distribution of $B\to \pi \ell \bar{\nu}_\ell$ in terms of the two kinematic variables $q^2$ and $\cos{\theta_{\ell}}$,
    \begin{align}
        \frac{d^2\Gamma(B\to \pi \ell \bar{\nu}_\ell)}{dq^2~d\cos{\theta_\ell}}=a_{\theta_\ell}(q^2)+b_{\theta_\ell}(q^2)\cos{\theta_\ell}+c_{\theta_\ell}(q^2)\cos^2{\theta_\ell},
        \label{eq:3.7}
    \end{align}
where the three coefficient functions are given by \cite{Becirevic:2016hea}
    \begin{align}
        &a_{\theta_\ell}(q^2)=\mathcal{N}_{\rm ew} \lambda^{3/2}\left(1-\frac{m_\ell^2}{q^2}\right)^2\left[\left| f_{B\pi}^+(q^2)\right|^2+\frac{1}{\lambda}\frac{m_\ell^2}{q^2}\left(1-\frac{m_\pi^2}{m_B^2}\right)^2\left| f_{B\pi}^0(q^2)\right|^2\right],\\
        &b_{\theta_\ell}(q^2)=2\mathcal{N}_{\rm ew} \lambda\left(1-\frac{m_\ell^2}{q^2}\right)^2\frac{m_\ell^2}{q^2}\left(1-\frac{m_\pi^2}{m_B^2}\right){\rm Re}\left[f_{B\pi}^+(q^2)f_{B\pi}^{0~*}(q^2)\right],\\
        &c_{\theta_\ell}(q^2)=-\mathcal{N}_{\rm ew}\lambda^{3/2}\left(1-\frac{m_\ell^2}{q^2}\right)^3\left| f_{B\pi}^+(q^2)\right|^2,
    \end{align}
with the symbols,
    \begin{align}
        &\mathcal{N}_{\rm ew}=\frac{G_F^2\left|V_{ub}\right|^2 m_B^3}{256\pi^3},~~~~\lambda\equiv \lambda\left(1,\frac{m_\pi^2}{m_B^2},\frac{q^2}{m_B^2}\right),\notag\\
        &\lambda(a,b,c)\equiv a^2+b^2+c^2-2ab-2ac-2bc.
    \end{align}
The helicity angle $\theta_{\ell}$ is defined as the angle between the momentum direction of $\ell^-$ lepton and the momentum direction of the final-state pion in the center-of-mass frame of the dilepton system.
It can be easily seen that in the massless lepton limit, the three coefficient functions satisfy a simple algebraic relation, i.e., $b_{\theta_\ell}(q^2)=0$ and $a_{\theta_\ell}(q^2)+c_{\theta_\ell}(q^2)=0$.

After integrating out the helicity angle $\theta_{\ell}$, we obtain the expression for the differential decay width of $B\to\pi\ell\bar{\nu}_\ell$ in the rest frame of the bottom meson
    \begin{align}
    \frac{d\Gamma(B\to\pi\ell\bar{\nu}_\ell)}{dq^2}
        =&\frac{4}{3}\mathcal{N}_{\rm ew}\lambda^{3/2}\left(1-\frac{m_\ell^2}{q^2}\right)^2\left\{\left(1+\frac{m_\ell^2}{2q^2}\right)\left|f_{B\pi}^+(q^2)\right|^2\right.\notag\\
        &~~~~\left.+\frac{1}{\lambda}\frac{3m_\ell^2}{2q^2}\left(1-\frac{m_\pi^2}{m_B^2}\right)^2\left|f_{B\pi}^0(q^2)\right|^2\right\}.
        \label{eq:branchingradio}
    \end{align}
In the massless limit $m_\ell=0$, we express Eq.(\ref{eq:branchingradio}) as
    \begin{align}
    \frac{d\Gamma(B\to\pi\ell\bar{\nu}_\ell)}{dq^2}
        =&\frac{4}{3}\mathcal{N}_{\rm ew}\lambda^{3/2}\left|f_{B\pi}^+(q^2)\right|^2,
    \end{align}
which means that the semileptonic decay branching ratio depends only on one form factor in this limit.

    \begin{figure}[th]
        \centering
            \includegraphics[width=\textwidth]{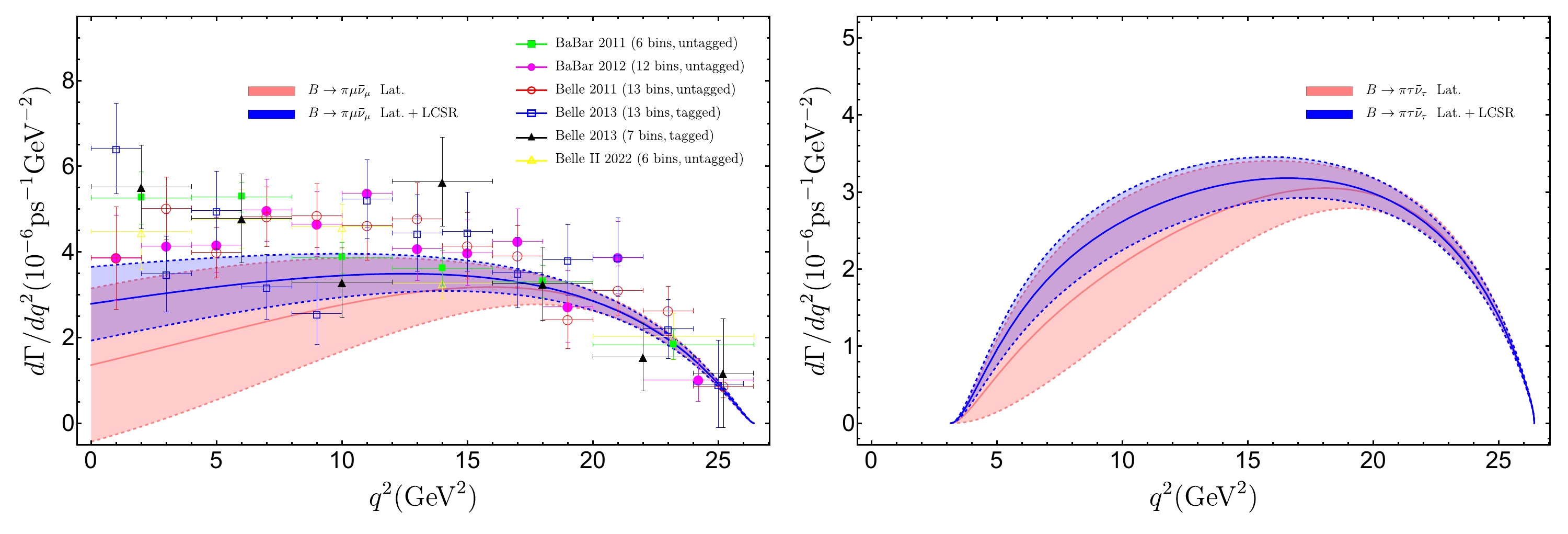}
        \vspace{-20pt}
        \captionsetup{labelfont=bf}
        \caption{ Theoretical predictions for the differential decay width of the   $B\to\pi\mu\bar{\nu}_\mu$   (left) and  $B\to\pi\tau\bar{\nu}_\tau$ (right) decay in the entire $q^2$ region, with or without LCSR as input. The   experimental measurements from the BaBar \cite{BaBar:2010efp,BaBar:2012thb}, Belle \cite{Belle:2010hep,Belle:2013hlo}, and Belle II \cite{Belle-II:2022imn} collaborations for the   decay   $B\to\pi\mu\bar{\nu}_\mu$ are displayed for a comparison.}
        \label{fig:Gammapic}
    \end{figure}

We present the differential $q^2$ distributions for the semileptonic decays $B\to\pi\mu\bar{\nu}_\mu$ and $B\to\pi\tau\bar{\nu}_\tau$ across the full kinematic range in Fig.\ref{fig:Gammapic}.
Numerical results for the BCL parameters are obtained from two fitting strategies: using lattice-QCD data alone, and combining lattice-QCD results with LCSR inputs.
The experimental data for the $B\to\pi\mu\bar{\nu}_\mu$ decay are taken from the BaBar \cite{BaBar:2010efp,BaBar:2012thb}, Belle \cite{Belle:2010hep,Belle:2013hlo}, and Belle II \cite{Belle-II:2022imn} collaborations.
A comparison with the theoretical predictions shows that the latter lie below the experimental data in the low-$q^2$ region.
Including the LCSR results with the $\mathcal{O}(\alpha_s^2\beta_0)$ contribution leads to a moderate upward shift of the theoretical prediction in the low-$q^2$ region.
With \textbf{PDG} recommended value $\left|V_{ub}\right|=(3.732^{+0.090}_{-0.085}) \times 10^{-3} $ \cite{ParticleDataGroup:2024cfk}, we obtain the branching fractions,
    \begin{align}
        &\mathcal{BR}(B\to\pi\mu\bar{\nu}_\mu)
  =(1.15^{+0.16}_{-0.15} )\times 10^{-4},\notag\\
        &\mathcal{BR}(B\to\pi\tau\bar{\nu}_\tau)
  =(0.82^{+0.08}_{-0.07})\times 10^{-4}.
    \label{eq:BR}
    \end{align}

Unlike the semileptonic decay $B\to\pi\mu\bar{\nu}_\mu$ measured by BaBar, Belle, and Belle II collaborations,
the semileptonic decay $B\to\pi\tau\bar{\nu}_\tau$ has not yet been observed, though Belle reports a 90$\%$ confidence level upper limit $\mathcal{B}(B\to\pi\tau\bar{\nu}_\tau) < 2.5 \times 10^{-4}$ \cite{Belle:2015qal}.
The $B\to\pi\tau\bar{\nu}_\tau$ decay is more challenging to measure because its final state contains at least two neutrinos, one from the $B$ decay and another from the subsequent $\tau$ decay, leading to significant missing energy.
Nevertheless, this channel is of particular interest due to its sensitivity to potential new-physics effects.
As a third-generation lepton, the $\tau$ is expected to be more sensitive to new physics. Prominent examples are the long-standing anomalies in the ratios $R_D$, $R_{D^*}$ \cite{Cheung:2020sbq}.
Motivated by these anomalies, several studies have proposed testing lepton flavor universality in the semileptonic  $B\to\pi\ell\bar{\nu}_\ell$ decays through the ratio $\mathcal{R}_\pi$,
\begin{align}
\mathcal{R}_\pi=\frac{\Gamma(B\to\pi\tau\bar{\nu}_\tau)}{\Gamma(B\to\pi\mu\bar{\nu}_\mu)}=\frac{\int^{q^2_{\rm max}}_{m_\tau^2}dq^2 \, {d\Gamma}(B\to\pi\tau\bar{\nu}_\tau)/dq^2}{\int^{q^2_{\rm max}}_{m_\mu^2}dq^2 \, {d\Gamma}(B\to\pi\mu\bar{\nu}_\mu)/dq^2}.
\end{align}
\begin{table}[thb]
        \footnotesize
    \centering\setlength\tabcolsep{6.0pt} \def\arraystretch{1.5}
                \caption{The lepton universality ratio $\mathcal{R}_\pi$ of the semileptonic $B\to\pi\ell\bar{\nu}_\ell$ decay calculated in the Standard Model, comparing with the LQCD results and the previous LCSR results.}
                \vspace{3pt}
        \begin{tabular}{|c|c|c|c|c|}
            \hline\hline
            Observables & LQCD & LCSR &LQCD $\oplus$ LCSR & This Work\\
            \hline\hline
                & & & $0.699\pm0.022$ \cite{Leljak:2021vte} &\\
            & $0.69\pm0.19$ \cite{Flynn:2015mha} & $0.69^{+0.03}_{-0.05}$ \cite{Khodjamirian:2011ub}& $0.677\pm0.010$ \cite{Biswas:2021cyd}&\\
            $\mathcal{R}_\pi$ & $0.767\pm0.145$ \cite{Martinelli:2022tte}  & $0.68^{+0.10}_{-0.09}$ \cite{Zhou:2019jny}& $0.78\pm0.10$ \cite{Becirevic:2020rzi} & $0.712\pm0.034$\\
            & $0.838\pm0.075$ \cite{Martinelli:2022tte} & $0.65^{+0.13}_{-0.11}$ \cite{Zhou:2019jny}& $0.720\pm0.027$ \cite{Cui:2022zwm} & \\
               & & & $0.746\pm0.039$ \cite{Cui:2022zwm}&\\
            \hline\hline
        \end{tabular}
        \label{tab:Rpi}
\end{table}

In Table~\ref{tab:Rpi}, we provide our prediction for $\mathcal{R}_\pi$ obtained from the combined analysis of lattice-QCD results at large $q^2$ and LCSR inputs including the $\mathcal{O}(\alpha_s^2\beta_0)$ contribution.
For comparison, we also show the lattice-QCD and previous LCSR results.
Our prediction is consistent with these existing results at the $1.0\sigma$ level.
Similarly, our prediction is consistent with the measurement $\mathcal{R}_\pi|_{\rm Belle~2016} = 1.05 \pm 0.51$ from the Belle Collaboration \cite{Belle:2015qal} at the $1.0\sigma$ level.

Unlike the total decay rates, which depend strongly on the form factors, the angular distributions of final states are less sensitive to nonperturbative hadronic parameters. These observables are therefore better suited for probing new physics signals with reduced hadronic uncertainties. We study the forward-backward asymmetry $\mathcal{A}_{\rm FB}$ and flat term $\mathcal{F}_{\rm H}$, which exhibit remarkable sensitivity to new physics associated with electroweak symmetry breaking \cite{Cui:2022zwm}
\begin{align}
        \mathcal{A}_{\rm FB}^{B\to\pi\ell\bar{\nu}_\ell}(q^2)&=\left[\frac{d\Gamma(B\to\pi\ell\bar{\nu}_\ell)}{dq^2}\right]^{-1}\int_{-1}^{1}d\cos{\theta_\ell}~{\rm sgn} (\cos{\theta_\ell})\frac{d^2\Gamma(B\to\pi\ell\bar{\nu}_\ell)}{dq^2d\cos{\theta_\ell}}\notag\\
        &=\left[\frac{1}{2}b_{\theta_\ell}(q^2)\right]:\left[a_{\theta_\ell}(q^2)+\frac{1}{3}c_{\theta_\ell}(q^2)\right],\\
        \mathcal{F}_{\rm H}^{B\to\pi\ell\bar{\nu}_\ell}(q^2)&=1+\frac{2}{3}\left[\frac{d\Gamma(B\to\pi\ell\bar{\nu}_\ell)}{dq^2}\right]^{-1}\frac{d^2}{d(\cos{\theta_\ell})^2}\frac{d^2\Gamma(B\to\pi\ell\bar{\nu}_\ell)}{dq^2d\cos{\theta_\ell}}\notag\\
        &=\left[a_{\theta_\ell}(q^2)+c_{\theta_\ell}(q^2)\right]:\left[a_{\theta_\ell}(q^2)+\frac{1}{3}c_{\theta_\ell}(q^2)\right].
\end{align}
It is evident that the two observables vanish in the massless lepton limit within the Standard Model. The polarization asymmetry of final-state leptons provides another relevant observable
\begin{align}
    \mathcal{A}_{\lambda_\ell}^{B\to\pi\ell\bar{\nu}_\ell}(q^2)&=\left[\frac{d\Gamma(B\to\pi\ell\bar{\nu}_\ell)}{dq^2}\right]^{-1}\left[\frac{d\Gamma^{\lambda_\ell=-1/2}}{dq^2}-\frac{d\Gamma^{\lambda_\ell=+1/2}}{dq^2}\right](B\to\pi\ell\bar{\nu}_\ell)\notag
 \\
    &=1-\frac{2}{3}\left\{\left[3a_{\theta_\ell}(q^2)+3c_{\theta_\ell}(q^2)+\frac{2m_\ell^2}{q^2-m_\ell^2}c_{\theta_\ell}(q^2)\right]:\left[a_{\theta_\ell}(q^2)+\frac{1}{3}c_{\theta_\ell}(q^2)\right]\right\},
\end{align}
which exhibits exceptional sensitivity to helicity-violating new physics interactions.

Using the numerical results obtained from the combined BCL $z$-series fit to lattice-QCD results and LCSR inputs including the $\mathcal{O}(\alpha_s^2\beta_0)$ contribution, we obtain the integrated predictions for $\mathcal{A}_{\rm FB}^{B\to\pi\ell\bar{\nu}_\ell}$, $\mathcal{F}_{\rm H}^{B\to\pi\ell\bar{\nu}_\ell}$, and $\mathcal{A}_{\lambda_\ell}^{B\to\pi\ell\bar{\nu}_\ell}$.
The results are collected in Table~\ref{tab:angular}, together with previous theoretical predictions for comparison.
They show good agreement at the $1.0\sigma$ level.
A slight discrepancy occurs in the $\mathcal{F}_{\rm H}^{B\to\pi\ell\bar{\nu}_\ell}$ prediction from \cite{Leljak:2021vte}, which is approximately one-fourth of our value. This was resolved in Ref.\cite{Cui:2022zwm} through additional kinematic constraints:
    \begin{align}
        \mathcal{F}_{\rm H}^{B\to\pi\mu\bar{\nu}_\mu}\geq6.337\times 10^{-4},~~~~~~~~\mathcal{F}_{\rm H}^{B\to\pi\tau\bar{\nu}_\tau}\geq0.169.
    \end{align}

    \begin{table}[htb]
        \footnotesize
        \centering\setlength\tabcolsep{6.0pt} \def\arraystretch{1.5}
        \caption{Theoretical predictions for the three  classes of integrated observables, $\mathcal{A}_{\rm FB}^{B\to\pi\ell\bar{\nu}_\ell}$, $\mathcal{F}_{\rm H}^{B\to\pi\ell\bar{\nu}_\ell}$, and $\mathcal{A}_{\lambda_\ell}^{B\to\pi\ell\bar{\nu}_\ell}$, obtained from the combined BCL $z$-series expansion fitting of the    lattice data   and our $B$-meson LCSR results for $B\to\pi$ form factors. Previous theoretical results are also shown for comparison.}
        \vspace{3pt}
        \begin{tabular}{|c|c|c|c|}
            \hline\hline
            Observables & LQCD  & LQCD $\oplus$ LCSR & This Work\\
            \hline\hline
                &&$(4.8\pm0.3)\times 10^{-3}$ \cite{Leljak:2021vte}&\\
            $\mathcal{A}_{\rm FB}^{B\to\pi\mu\bar{\nu}_\mu}$ & $(4.4\pm1.3)\times 10^{-3}$ \cite{Flynn:2015mha}   & $(3.99\pm0.35)\times10^{-3}$ \cite{Cui:2022zwm}& {$(4.10\pm0.48)\times10^{-3}$}  \\
            &    & $(3.72\pm0.51)\times10^{-3}$ \cite{Cui:2022zwm} & \\
            \hline\hline
            &    & $0.259\pm0.004$ \cite{Leljak:2021vte} &\\
            $\mathcal{A}_{\rm FB}^{B\to\pi\tau\bar{\nu}_\tau}$ & $0.252\pm0.012$ \cite{Flynn:2015mha} & $0.248\pm0.005$\cite{Cui:2022zwm}& {$0.249\pm 0.005 $}  \\
            &    & $0.244\pm0.007$ \cite{Cui:2022zwm} &\\
            \hline\hline
            &    &$(2.4\pm0.1)\times10^{-3}$ \cite{Leljak:2021vte} &\\
            $\mathcal{F}_{\rm H}^{B\to\pi\mu\bar{\nu}_\mu}$   & --  & $(8.04\pm0.72)\times10^{-3}$ \cite{Cui:2022zwm} & {$(8.25\pm0.96)\times10^{-3}$} \\
            &    & $(7.52\pm1.02)\times10^{-3}$ \cite{Cui:2022zwm} & \\
            \hline\hline
            &     & $0.134\pm0.003$ \cite{Leljak:2021vte} & \\
            $\mathcal{F}_{\rm H}^{B\to\pi\tau\bar{\nu}_\tau}$ & --  & $0.514\pm0.012$ \cite{Cui:2022zwm}  & {$0.515\pm0.010$} \\
            &    &$0.508\pm0.014$ \cite{Cui:2022zwm} & \\
            \hline\hline
            &   & $0.988\pm0.001$ \cite{Cui:2022zwm} &\\
            $\mathcal{A}_{\lambda_\ell}^{B\to\pi\mu\bar{\nu}_\mu}$ & --  &  & {$0.989\pm0.001$ }  \\
            &    & $0.989\pm0.002$ \cite{Cui:2022zwm} &  \\
            \hline\hline
            &    & $0.21\pm0.02$ \cite{Leljak:2021vte}  &  \\
            $\mathcal{A}_{\lambda_\ell}^{B\to\pi\tau\bar{\nu}_\tau}$ & --   & $0.266\pm0.029$ \cite{Cui:2022zwm} & {$0.267\pm0.025$ }  \\
            &    & $0.272\pm0.032$ \cite{Cui:2022zwm}&\\
            \hline\hline
        \end{tabular}
        \label{tab:angular}
    \end{table}

\subsection{Extraction of \texorpdfstring{$V_{ub}$}{Vub}}

The semileptonic decay channel $B \to \pi \mu \bar{\nu}_\mu$ is commonly used to extract the CKM matrix element $\vert V_{ub} \vert$.
The inclusion of the $\mathcal{O}(\alpha_s^2\beta_0)$ contribution motivates an updated determination of $\vert V_{ub} \vert$.
This can be achieved by fitting the experimental branching-fraction data together with the form-factor constraints from LCSR and LQCD over the full kinematic region.
Using the available experimental data, including three untagged datasets from BaBar \cite{BaBar:2010efp,BaBar:2012thb} and Belle \cite{Belle:2010hep}, two tagged datasets for $\bar{B}^0\to\pi^+\mu\bar{\nu}_\mu$ and $B^-\to\pi^0\mu\bar{\nu}_\mu$ from Belle \cite{Belle:2013hlo}, and an untagged dataset for $B^0\to\pi^-\mu\bar{\nu}_\mu$ from Belle II \cite{Belle-II:2022imn}, we perform a $q^2$-binned minimal-$\chi^2$ fit.
There are five free coefficients in the BCL parametrization and one additional free parameter $\left|V_{ub}\right|$.
The goodness of fit is quantified by $\chi_{\text{min}}^2/{\rm d.o.f.} = 77.15/(74-6) = 1.13$.
Table~\ref{tab:BCL parameters and vub} presents the central values, uncertainties, and correlation matrix for the fitted BCL $z$-expansion parameters and $\left|V_{ub}\right|$, with the $z$-series expansion truncated at $N=3$.

The BCL parameters obtained from the joint fit to experimental data are consistent at the $1.0\sigma$ level with those in Table~\ref{tab:BCL parameters}, which are fitted from the combined LCSR and LQCD results for the form factors.
In addition, the value of $\left|V_{ub}\right|$ from our global fit, shown in Table~\ref{tab:BCL parameters and vub}, is consistent with the LCSR result based on pion LCDAs \cite{Leljak:2021vte},
\begin{align}
\left|V_{ub}\right|_{B\to\pi\ell\bar{\nu}_\ell} = (3.77 \pm 0.15) \times 10^{-3} ,
\end{align}
and with the BCL fit using LQCD results only,
\begin{align}
\left|V_{ub}\right|_{B\to\pi\ell\bar{\nu}_\ell} = (3.66 \pm 0.15) \times 10^{-3} .
\end{align}
The uncertainty is comparable to those of the previous determinations.

    \begin{table}[th]
        \footnotesize
        \centering\setlength\tabcolsep{6.0pt} \def\arraystretch{1.5}
        \caption{The central values, errors, and correlation matrix of the $z$-series coefficients for the form factors and the CKM matrix element $\left|V_{ub}\right|$ in the BCL parametrization, obtained from the combined $\chi^2$ fit to LCSR, LQCD \cite{Flynn:2015mha,Colquhoun:2022atw,FermilabLattice:2015mwy}, and experimental data points \cite{BaBar:2010efp,BaBar:2012thb,Belle:2010hep,Belle:2013hlo,Belle-II:2022imn} for $N=3$.}
        \vspace{3pt}
        \begin{tabular}{|c|c||cccccc|}
            \hline\hline
            \multicolumn{2}{|c||}{$B\to \pi$ Form Factors} \vline& \multicolumn{6}{|c|}{Correlation Matrix}  \\
            \hline\hline
            Parameters & Values & $\left|V_{ub}\right|$ & $b_0^+$ & $b_1^+$ & $b_2^+$ & $b_0^0$ & $b_1^0$ \\
            \hline\hline
            $\left|V_{ub}\right|$ & $3.73(14)\times10^{-3}$ & 1 & -0.829 & -0.381 & 0.373 & -0.071 & -0.510\\
            $b_0^+$ & 0.405(12) & & 1 & 0.189 & -0.476 & 0.053 & 0.416 \\
            $b_1^+$ & -0.516(46) & &  & 1 & -0.768 & -0.027 & 0.234 \\
            $b_2^+$ & -0.124(167) & &  &  & 1 & 0.060 & -0.094 \\
            $b_0^0$ & 0.497(7) & &  &  &  & 1 & 0.460 \\
            $b_1^0$ & -1.414(37) & &  &  &  &  & 1 \\
            \hline\hline
        \end{tabular}
        \label{tab:BCL parameters and vub}
    \end{table}

\section{Conclusion}

In this work, we have computed, for the first time, the $\mathcal{O}(\alpha_s^2\beta_0)$ QCD correction to the $B\to\pi$ form factors within the LCSR approach based on $B$-meson light-cone distribution amplitudes.

Using the method of regions, we analyzed the factorization formula of the correlation function, in which the short-distance coefficients are factorized into hard and jet functions.
We computed the jet function at $\mathcal{O}(\alpha_s^2\beta_0)$ by inserting an effective dressed gluon propagator in the large-$\beta_0$ limit.
This provides a definite and calculable two-loop contribution to the hard-collinear kernel and isolates the running-coupling-induced part of the NNLO correction in the LCSR framework.
We emphasize that this result should be interpreted as a partial NNLO correction, rather than as a complete NNLO prediction for the physical form factors.
The full NNLO result may receive additional contributions from non-$\beta_0$-enhanced two-loop topologies and radiative corrections involving three-particle $B$-meson LCDAs.
Accordingly, the relatively small numerical impact of the $\mathcal{O}(\alpha_s^2\beta_0)$ term should not be over-interpreted as evidence that all missing NNLO effects are small.
Nevertheless, the present calculation provides a useful benchmark for the perturbative structure of the LCSR and can serve as a first step toward a systematic analysis of renormalon effects in the $B\to\pi$ form factors.

Additionally, we calculated the UV-divergent terms of the $B$-meson light-cone distribution amplitudes, thereby verifying the factorization scale independence at $\mathcal{O}(\alpha_s^2\beta_0)$. {
Neglecting the three-particle $B$-meson LCDAs, we also check that the $\mathcal{O}(\alpha_s^2\beta_0)$ ultraviolet structures of $\phi_B^+(\omega)$ and $\phi_B^-(\omega)$ are consistent with the relation
implied by the Wandzura-Wilczek approximation.}

Our $\mathcal{O}(\alpha_s^2\beta_0)$ QCD corrections to the $B\to\pi$  form factors amounted to approximately $+6.1\%$ relative to the leading-order prediction.
Additionally, we combined the numerical results of the form factors obtained from LCSR in the large recoil region with LQCD results in the small recoil region to perform a global fit. Employing the BCL $z$-expansion, we obtained the $q^2$ dependence of   $B\to\pi$ transition form factors in the entire kinematic range.
Utilizing the fitted form factors, we provided numerical results for the branching ratios of $B\to\pi\mu\bar{\nu}_{\mu}$ and $B\to\pi\tau\bar{\nu}_{\tau}$ decays. We also give numerical results of forward-backward asymmetry, $\mathcal{A}_{\mathrm{FB}}^{B\to\pi\ell\bar{\nu}_{\ell}}$, flat term $\mathcal{F}_{\mathrm{H}}^{B\to\pi\ell\bar{\nu}_{\ell}}$, and polarization asymmetry $\mathcal{A}_{\lambda_\ell}^{B\to\pi\ell\bar{\nu}_{\ell}}$ with smaller theoretical uncertainty, which are sensitive to new physics signals beyond the Standard Model.
We also predicted the lepton flavor universality ratio $\mathcal{R}_{\pi}$ for exclusive $B\to\pi\ell\bar{\nu}_{\ell}$ decays:
\begin{align}
\mathcal{R}_{\pi} = 0.712 \pm 0.034 \qquad \text{(BCL fit with truncation order $N=3$)}.
\end{align}
Finally, we extracted the CKM matrix element $|V_{ub}|$ by combining the fitted form factors with experimental data on $B\to\pi\mu\bar{\nu}_{\mu}$ from BaBar, Belle, and Belle II collaborations:
\begin{align}
|V_{ub}| = (3.73 \pm 0.14) \times 10^{-3} \qquad \text{(BCL fit with truncation order $N=3$)}.
\end{align}

{
Future improvements of the heavy-to-light form factors can be pursued along several directions.
First, performing the full two-loop QCD calculation for the $B\to\pi$ form factors would be numerically interesting for assessing the reliability of the perturbative expansion and for determining whether
the $\mathcal{O}(\alpha_s^2\beta_0)$ contribution provides a reasonable estimate of the complete two-loop correction.
}
Second, investigating the next-to-leading-power corrections at the $\mathcal{O}(\alpha_s)$ level would be important for understanding the factorization properties of the exclusive semileptonic $B$-meson decays.
Third, the strategies discussed in this work can be extended to analogous heavy-to-light $B_{(s)} \to \rho,\omega,\phi,K^{(*)}$ form factors.
Fourth, improving the unitarity bounds for the $z$-series parametrization of $B\to \pi$ form factors will be helpful in constraining their momentum-transfer dependence.
Fifth, advancing our knowledge of the $B$-meson LCDA, from LQCD calculations \cite{Han:2024min,LatticeParton:2024zko} or global fits \cite{Li:2025mhq}, will be indispensable for reducing the uncertainties of theoretical predictions.

\section*{Acknowledgement}

We are pleased to acknowledge Yu-Ming Wang for enlightening discussions and to Lei-Yi Li for collaboration in the early stages of this project.
The work of Y.K.H., B.X.S. and H.X.Y. is supported by the National Natural Science Foundation of China under grants No. 12475097, 12535006 and the Natural Science Foundation of Tianjin with grant No. 25JCZDJC01190, { and by the Fundamental Research Funds for the Central Universities with grant No. 63261180.}
Dong-Hao Li acknowledges support from the National Natural Science Foundation of China under grant No. 12447154.
Cai-Dian L\"u acknowledges support from the National Key Research and Development Program of China (2023YFA1606000), National Natural Science Foundation of China under grants No. 12275277, 12435004.

\appendix

\section{Soft Subtraction and Factorization-scale Independence}\label{app-softsub}
     \begin{figure}[htb]
        \centering
        \begin{subfigure}[t]{0.25\textwidth}
            \includegraphics[width=\textwidth]{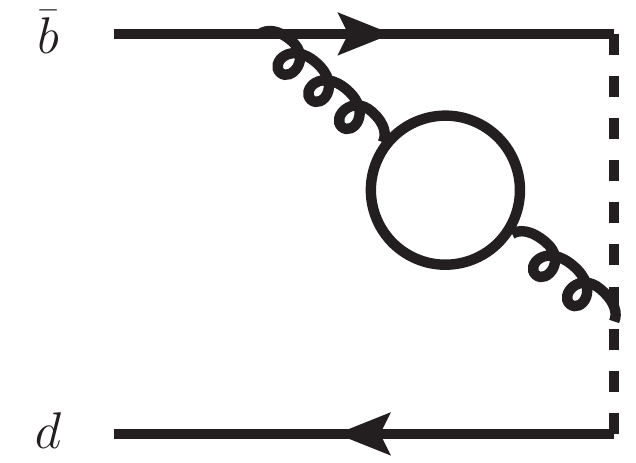}
            \caption{}
            \label{fig:lcda_a_two}
        \end{subfigure}
        \hspace{0.05\textwidth}
        \begin{subfigure}[t]{0.25\textwidth}
            \includegraphics[width=\textwidth]{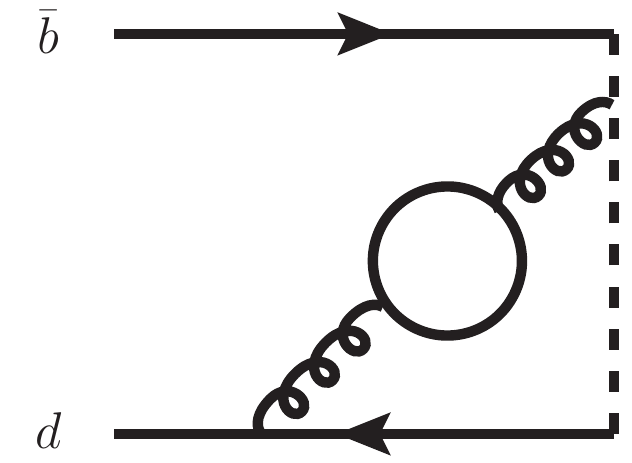}
            \caption{}
            \label{fig:lcda_b_two}
        \end{subfigure}
        \hspace{0.05\textwidth}
        \begin{subfigure}[t]{0.25\textwidth}
            \includegraphics[width=\textwidth]{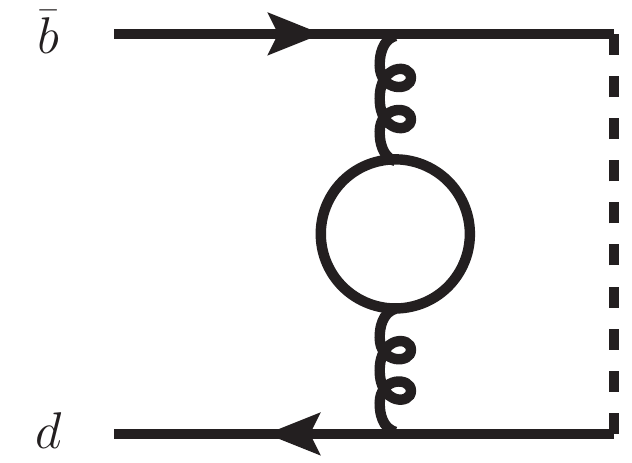}
            \caption{}
            \label{fig:lcda_c_two}
        \end{subfigure}
        \captionsetup{labelfont=bf}
        \caption{Two-loop corrections for the partonic distribution amplitude $\Phi_{b\bar{d}}(\omega')$.}
        \label{fig:lcda_two}
    \end{figure}

In this appendix, we present the soft subtraction and factorization-scale analysis at $\mathcal{O}(\alpha_s^2\beta_0)$ as an important internal check on the correctness of our two-loop diagrammatic calculation.

After introducing the effective dressed gluon propagator $\mathcal{D}^{ab}_{\mu\nu}(\ell)$ in Eq.~\eqref{eq:dressed-gluon-propagator}, the pNNLO diagrams in the large-$\beta_0$ limit can be regarded as the corresponding NLO graphs with the gluon propagator replaced by the building block $\mathcal{D}^{ab}_{\mu\nu}(\ell)$. Therefore, the factorization analysis at $\mathcal{O}(\alpha_s^2\beta_0)$ follows the same pattern as at NLO: the soft contribution to the partonic correlation function is cancelled by the perturbative correction to the partonic distribution amplitude, while the hard-collinear remainder defines the jet function.
In Eq.~\eqref{eq:2loophctot}, the jet function $J^{(2)}$ is extracted from the finite part of the hard-collinear contribution $\Pi_{\mu}^{(2),hc}$. The corresponding matching condition relies on the cancellation between the soft-region contribution $\Pi_{\mu}^{(2),s}$ and the infrared subtraction term $\Phi_{b\bar{d}}^{(2)} \otimes T^{(0)}$, in analogy with the NLO analysis.

Applying the method of regions and assuming the loop momentum  scales as $ \ell^\mu  \sim (\Lambda,\Lambda,\Lambda)$, we obtain the soft contribution of $\Pi_{\mu}^{(2a)}(p_B,q)$
     \begin{align}\label{eqappA1}
         \Pi_{\mu}^{(2a),s}(p_B,q)
        =&\frac{g_s^4 C_F N_f T_F}{\bar{n}\cdot p-\omega}\frac{4i(-1)^{1-\epsilon}}{(4\pi)^{2-\epsilon}}\frac{\Gamma(\epsilon)\Gamma^2(2-\epsilon)}{\Gamma(4-2\epsilon)}
  \times \bar{d}(k)\slashed{n}\gamma_5 \slashed{\bar{n}}\gamma_\mu m_b(1+\slashed{v})\left(1-\frac{\bar{n}\cdot \ell}{\ell^2}v\cdot \ell\right)b(v)
  \notag \\
  &\times\mu^{4\epsilon}\int\frac{d^D \ell}{(2\pi)^D}\frac{1}{\left(\bar{n}\cdot p-\omega+\bar{n}\cdot \ell\right)2m_b v\cdot \ell(\ell^2)^{1+\epsilon}}.
     \end{align}
With the Wilson-line Feynman rules, the corresponding subtraction term in Fig.\ref{fig:lcda_a_two} reads:
\begin{align}\label{eqappA2}
        \Phi_{b\bar{d}}^{(2a)}(\omega,\omega')=
        &g_s^4 C_F N_f T_F \frac{8i(-1)^{1-\epsilon}}{(4\pi)^{2-\epsilon}}\frac{\Gamma(\epsilon )\Gamma^2(2-\epsilon )}{\Gamma(4-2\epsilon)}
        \times \left[\bar{d}(k)\right]_\alpha\left[(m_b\slashed{v}+m_b)\left(1-\frac{\bar{n}\cdot \ell}{\ell^2}v\cdot \ell\right)b(v)\right]_\beta
        \notag\\
        &\times\mu^{4\epsilon} \int \frac{d^D\ell}{(2\pi)^D}\frac{\delta(\omega'-\omega+\bar{n}\cdot \ell)-\delta(\omega'-\omega)}{(\ell^2)^{1+\epsilon}\left[2m_b v\cdot \ell \right]\bar{n}\cdot \ell}.
\end{align}

From Eq.(\ref{eqappA1},\ref{eqappA2}) one can deduce the absence of soft contribution to the perturbative matching coefficient at $\mathcal{O}(\alpha_s^2\beta_0)$
     \begin{align}
        \Phi_{b\bar{d}}^{(2a)} \otimes  T^{(0)}=\Pi_{\mu}^{(2a),s}
     \end{align}
with the tree level hard-scattering kernel given by,
\begin{align}
    T^{(0)}(n\cdot p , \bar{n} \cdot p, \omega') =  \frac{1}{2}\frac{1}{\bar{n}\cdot p - \omega' +i0} \left[ \slashed{n}\gamma_5 \slashed{\bar{n}} \gamma_\mu \right]_{\alpha\beta}.
\end{align}
Following the same procedure as before, we present the soft region contributions of $\Pi_{\mu}^{(2)}(p_B,q)$ in Figs.\ref{fig:pion} and \ref{fig:box},
 \begin{align}
     \Pi_{\mu,}^{(2b),s}(p_B,q)
     =&-\frac{g_s^4 C_F N_f T_F}{\bar{n}\cdot p -\omega}\frac{8i(-1)^{1-\epsilon}}{(4\pi)^{2-\epsilon}}\frac{\Gamma(\epsilon)\Gamma^2(2-\epsilon)}{\Gamma(4-2\epsilon)}{\bar{n}_\mu}\notag\\
     &\times \bar{d}(k)\left[\slashed{\bar{n}}(\slashed{k}_\perp+\slashed{\ell}_\perp)-\frac{(\ell+k)^2}{\ell^2}\bar{n}\cdot \ell\right]\slashed{n}\gamma_5 b(v)
     \notag\\
     &\times \mu^{4\epsilon}\int \frac{d^D\ell}{(2\pi)^D}\frac{1}{(\ell^2)^{1+\epsilon}(\ell+k)^2\left[\bar{n}\cdot p-\omega-\bar{n}\cdot \ell\right]}.
 \end{align}
 and
 \begin{align}
      \Pi_{\mu}^{(2d),s}(p_B,q)=&g_s^4 C_F N_f T_F\frac{8i(-1)^{-\epsilon}}{(4\pi)^{2-\epsilon}}\frac{\Gamma(\epsilon)\Gamma^2(2-\epsilon)}{\Gamma(4-2\epsilon)}{\bar{n}_\mu}
      \times \left\{
      \mu^{4\epsilon}\int\frac{d^D\ell}{(2\pi)^D}\frac{\bar{d}(k)\slashed{n}\gamma_5 b(p_B-k)}{(\ell^2)^{2+\epsilon}[\bar{n}\cdot \ell-(\omega-\bar{n}\cdot p)]}
      \right. \notag \\
      &~\left.+\mu^{4\epsilon}\int\frac{d^D\ell}{(2\pi)^D}\frac{\bar{d}(k)\slashed{v}(\slashed{k}-\slashed{\ell})\slashed{n}\gamma_5 b(p_B-k)}{(\ell^2)^{1+\epsilon}(\ell-k)^2 v\cdot \ell [\bar{n}\cdot \ell-(\omega-\bar{n}\cdot p)] }
      \right\}.
\end{align}
The soft region of $\Pi_{\mu}^{(2c)} =  0$ in the leading power of $\Lambda_{\text{QCD}}/m_b$.
The corresponding inferred subtraction terms in Figs.\ref{fig:lcda_b_two} and \ref{fig:lcda_c_two} read:
\begin{align}
    \Phi_{b\bar{d}}^{(2b)}(\omega,\omega')=&- g_s^4 C_F N_f T_F \frac{8i(-1)^{1-\epsilon}}{(4\pi)^{2-\epsilon}}\frac{\Gamma(\epsilon )\Gamma^2(2-\epsilon )}{\Gamma(4-2\epsilon)}
    \times \left[\bar{d}(k)\left(\slashed{\bar{n}}-\frac{\bar{n}\cdot \ell}{\ell^2}\slashed{\ell}\right)\left(\slashed{\ell}+\slashed{k}\right)\right]_\alpha\left[b(v)\right]_\beta
    \notag\\
        &\times\mu^{4\epsilon} \int \frac{d^D\ell}{(2\pi)^D}\frac{\delta(\omega'-\omega)-\delta(\omega'-\omega-\bar{n}\cdot \ell)}{(\ell^2)^{1+\epsilon}(\ell+k)^2\bar{n}\cdot \ell},
\end{align}
and
\begin{align}
      \Phi_{b\bar{d}}^{(2c)}(\omega,\omega')=&-g_s^4C_FN_fT_F\frac{8i(-1)^{1-\epsilon}}{(4\pi)^{2-\epsilon}}\frac{\Gamma(\epsilon)\Gamma^2(2-\epsilon)}{\Gamma(4-2\epsilon)}
      \times \left[\bar{d}(k)\gamma^\rho(\slashed{\ell}-\slashed{k})\right]_{\alpha}\left[(\slashed{\ell}+m_b\slashed{v}+m_b)\gamma^{\rho'}b_v\right]_\beta \notag \\
      &\times \mu^{4\epsilon}\int\frac{d^D\ell}{(2\pi)^D}\frac{\delta(\omega'-\omega+\bar{n}\cdot \ell)}{(\ell^2)^{1+\epsilon}(\ell-k)^2v\cdot \ell}\cdot \frac{1}{2m_b}\left(
      g_{\rho\rho'}-\frac{\ell_\rho \ell_{\rho'}}{\ell^2}\right).
\end{align}
  After convoluting the partonic DA with the tree-level hard-scattering kernel, it can be seen that the infrared subtraction term cancels the soft contribution of the correlation function completely.

Moreover, we regularize infrared divergences using a mass $m$ and compute the ultraviolet divergent (UV) terms in $\Phi_{b\bar{d}}$. Upon summing the UV divergences from the hard function, the jet function, and $\Phi_{b\bar{d}}$, we find that they cancel exactly. This cancellation verifies the factorization-scale independence of the correlation function at $\mathcal{O}(\alpha_s^2 \beta_0)$ accuracy, as required by QCD factorization:
\begin{align}
    \frac{d}{d \ln \mu} \Pi_{n,\bar{n}}(n\cdot p,\bar{n}\cdot p)=0
\end{align}
We present our calculation of the UV divergent terms in $\Phi_{b\bar{d}}$ and project them onto $\phi_{b\bar{d}}^+$ and $\phi_{b\bar{d}}^-$ using the light-cone projector in momentum space \cite{DeFazio:2007hw}
\begin{align}
M_{\beta \alpha}=  -\frac{i \tilde{f}_{B}(\mu) m_{B}}{4} \times\left\{\frac{1+\slashed{v}}{2}\left[\phi_{b\bar{d}}^{+}\left(\omega^{\prime}\right)\slashed{n}+
\phi_{b\bar{d}}^{-}\left(\omega^{\prime}\right) \slashed{\bar{n}}-\frac{2 \omega^{\prime}}{D-2} \phi_{b\bar{d}}^{-}\left(\omega^{\prime}\right) \gamma_{\perp}^{\rho} \frac{\partial}{\partial k_{\perp \rho}^{\prime}}\right] \gamma_{5}\right\}_{\alpha \beta}
\end{align}
The UV divergences of $\tilde{f}_B \phi_{b\bar{d}}^{\pm}(\omega')$ at $\mathcal{O}(\alpha_s^2)$ in the large-$\beta_0$ limit are collected below,
\begin{align}\label{appeq:phip2}
        \left[\tilde{f}_B\phi_{b\bar{d}}^+(\omega')\right]^{(2)}&= \left ( \frac{\alpha_s}{4\pi} \right )^2\tilde{f}_B C_F T_F N_f \notag\\
        &\times\left\{ \phi_{b\bar{d}}^+(\omega') \left[\frac{1}{\epsilon^3} +\frac{1}{9\epsilon^2}(12\ln \frac{\mu}{\omega'}-20) +\frac{1}{54\epsilon}(13+15\pi^2-120\ln \frac{\mu}{\omega'})\right]   \right.\notag \\
        &~~~~+\int_0^\infty d\omega \left[\frac{\theta(\omega'-\omega)}{\omega-\omega'}-\frac{\omega'\theta(\omega-\omega')}{\omega(\omega-\omega')}\right]_{\oplus}\phi_{b\bar{d}}^+(\omega)\left[\frac{4}{3\epsilon^2}-\frac{20}{9\epsilon}\right]\notag\\
        &\left.~~~~-\int_0^\infty d\omega~ \frac{\omega'\theta(\omega-\omega')}{\omega(\omega-\omega')}\ln{\left(\frac{\omega}{\omega'}\right)}\phi_{b\bar{d}}^+(\omega)\frac{4}{3\epsilon}\right\}.
 \end{align}
 \begin{align}\label{appeq:phim2}
        \left[\tilde{f}_B\phi_{b\bar{d}}^-(\omega')\right]^{(2)}&= \left ( \frac{\alpha_s}{4\pi} \right )^2\tilde{f}_B C_F T_F N_f \notag\\
        &\times\left\{ \phi_{b\bar{d}}^-(\omega') \left[\frac{1}{\epsilon^3} +\frac{1}{9\epsilon^2}(12\ln \frac{\mu}{\omega'}-20) +\frac{1}{54\epsilon}(13+15\pi^2-120\ln \frac{\mu}{\omega'})\right]   \right.\notag \\
        &~~~~+\int_0^\infty d\omega \left[\frac{\theta(\omega'-\omega)}{\omega-\omega'}-\frac{\omega'\theta(\omega-\omega')}{\omega(\omega-\omega')}\right]_{\oplus}\phi_{b\bar{d}}^-(\omega)\left[\frac{4}{3\epsilon^2}-\frac{20}{9\epsilon}\right]\notag\\
        &~~~~-\int_0^\infty d\omega ~\frac{\omega'\theta(\omega-\omega')}{\omega(\omega-\omega')}\ln{\left(\frac{\omega}{\omega'}\right)} \phi_{b\bar{d}}^-(\omega)\frac{4}{3\epsilon}\notag\\
        &\left.~~~~-\int_0^\infty d \omega~ \theta (\omega-\omega') \frac{\phi_{b\bar{d}}^-(\omega)}{\omega}\left[\frac{4}{3\epsilon^2}-\frac{1}{9\epsilon}(20+12\ln \frac{\omega}{\omega'})\right]\right\}.
\end{align}{
The RGE of the HQET decay constant $\tilde f_B$ is known to two-loop accuracy
\cite{Broadhurst:1994se}. The evolution equation of the leading-twist
$B$-meson LCDA $\phi_B^+(\omega)$ was derived at one loop in
Ref.~\cite{Lange:2003ff} and at two loops in Ref.~\cite{Braun:2019wyx}. We use the momentum-space representation summarized in
Ref.~\cite{Galda:2020epp},}
    \begin{align}
        \frac{d}{d\ln \mu}  \tilde{f}_B(\mu)&=\tilde{\gamma}(\alpha_s)\tilde{f}_B(\mu), \\
     \frac{d }{d \ln \mu} \phi^{+}_{B}(\omega', \mu) &=
     \left[-\Gamma_{\rm c}(\alpha_s) \ln \frac{\mu}{\omega'}-\gamma_\eta(\alpha_s) \right]\phi_{B}^+(\omega',\mu)
     \notag\\
        &+\Gamma_{c}(\alpha_s)\int_0^\infty d\omega \left[\frac{\omega'\theta(\omega-\omega')}{\omega(\omega-\omega')} +\frac{\theta(\omega'-\omega)}{\omega'-\omega} \right]_\oplus\phi_{B}^+(\omega,\mu)
  \notag\\
        &+\int_0^\infty d\omega \, \hat{\gamma}_+(\omega',\omega;\alpha_s)
\phi_{B}^+(\omega,\mu).
    \end{align}
with the anomalous dimension are expanded as
 \begin{align}
 \tilde{\gamma}\left(\alpha_{s}\right) & =\frac{\alpha_{s} C_{F}}{4 \pi}\left[\tilde{\gamma}^{(0)}+\left(\frac{\alpha_{s}}{4 \pi}\right) \tilde{\gamma}^{(1)}+\ldots\right], \notag
 \\
  \gamma_{\eta}\left(\alpha_{s}\right) & =\frac{\alpha_{s} C_{F}}{4 \pi}\left[\gamma_\eta^{(0)}+\left(\frac{\alpha_{s}}{4 \pi}\right) \gamma_\eta^{(1)}+\ldots\right], \notag
 \\
\Gamma_{\mathrm{c}}\left(\alpha_{s}\right)&=\frac{\alpha_{s}C_F}{4 \pi}\left[\Gamma^{(0)}_c +\left(\frac{\alpha_{s}}{4 \pi}\right) \Gamma^{(1)}_c+\ldots\right],
 \notag \\
 \hat{\gamma}_+(\omega',\omega;\alpha_s)
&=C_F\left(\frac{\alpha_s}{2\pi}\right)^2
\frac{\omega'\theta(\omega-\omega')}{\omega(\omega-\omega')}\,
h\left(\frac{\omega'}{\omega}\right) +\ldots,
 \end{align}
 and
\begin{align}
     \tilde{\gamma}^{(0)} & =3, \quad \tilde{\gamma}^{(1)}=\frac{127}{6}+\frac{14 \pi^{2}}{9}-\frac{5}{3} N_{f}, \notag \\
\gamma_\eta^{(0)}&=-2,\quad \Gamma_0=4,\quad  \Gamma_1=4 \left[\left(\frac{67}{9}-\frac{\pi^{2}}{3}\right) C_{A}-\frac{20}{9} T_{F} N_{f}\right] , \notag \\
\gamma_\eta^{(1)}&= C_{F}\left(-4+\frac{14 \pi^{2}}{3}-24 \zeta_{3}\right)+C_{A}\left(\frac{254}{27}-\frac{55 \pi^{2}}{18}-6 \zeta_{3}\right)+T_{F} N_{f}\left(-\frac{64}{27}+\frac{10 \pi^{2}}{9}\right), \notag \\
h(x)&=\ln x\left\{\beta_0+2C_F\left[
\ln x-\frac{1+x}{x}\ln(1-x)-\frac{3}{2}\right]\right\}.
\end{align}
The coefficients proportional to $N_f$ in the $1/\epsilon$ pole of Eq.~\eqref{appeq:phip2} agree with the corresponding $N_f$-dependent
terms of the two-loop anomalous dimensions given above. This agreement provides a consistency check of the $\mathcal{O}(\alpha_s^2\beta_0)$ ultraviolet structure obtained in our calculation.

\section{The relation between the RGEs of \texorpdfstring{$\phi_B^+(\omega)$}{phiB+} and twist-three \texorpdfstring{$\phi_B^-(\omega)$}{phiB-} in the Wandzura-Wilczek approximation}\label{appwwapprox}

In the previous appendix, we verified that the UV-divergent term of $\phi_B^+(\omega)$ in the large-$\beta_0$ limit agrees with the existing result in the literature. Since the two-loop ultraviolet divergence of $\phi_B^-(\omega)$ is still not available at present, in this appendix we use the Wandzura-Wilczek (W-W) relation to check the relation between $\phi_B^+(\omega)$ and $\phi_B^-(\omega)$. We emphasize that the W-W approximation amounts to neglecting the three-particle $B$-meson LCDAs. A genuinely non-trivial analysis would require the inclusion of the three-particle contributions, which is beyond the scope of the present work.

{The equation of motion gives the standard W-W relation between $\phi_B^+(\omega)$ and $\phi_B^-(\omega)$ \cite{Beneke:2000wa}.
We use its $D$-dimensional generalization derived in
Ref.~\cite{DeFazio:2007hw}},
\begin{align}\label{appWWd}
        \omega\frac{d}{d\omega}\phi_B^-(\omega)=-\phi_B^+(\omega)+\frac{4-D}{2}\left[\phi_B^+(\omega)-\phi_B^-(\omega)\right].
\end{align}
In Ref.~\cite{Bell:2008er}, it was shown that the UV divergent part of $\phi_B^-(\omega)$ can be inferred from that of $\phi_B^+(\omega)$ through the W-W relation, leading to the following relation between the one-loop anomalous dimension kernels:
\begin{align}
    \gamma_{-}^{(0)}\left(\omega, \omega^{\prime} ; \mu\right)=\gamma_{+}^{(0)}\left(\omega, \omega^{\prime} ; \mu\right)-\Gamma_{\mathrm{c}}^{(0)} \frac{\theta\left(\omega^{\prime}-\omega\right)}{\omega^{\prime}},
\end{align}
where the anomalous dimension kernels are defined as,
\begin{align}
    \frac{d}{d \ln \mu} \phi_{B}^{+}(\omega ; \mu)
    &=
    -\frac{\alpha_{s} C_{F}}{4 \pi} \int_{0}^{\infty} d \omega^{\prime} \gamma_{+}^{(0)}\left(\omega, \omega^{\prime} ; \mu\right) \phi_{B}^{+}\left(\omega^{\prime} ; \mu\right)+\mathcal{O}\left(\alpha_{s}^{2}\right),
    \notag \\
    \frac{d}{d \ln \mu} \phi_{B}^{-}(\omega ; \mu)
    &=
    -\frac{\alpha_{s} C_{F}}{4 \pi} \int_{0}^{\infty} d \omega^{\prime} \gamma_{-}^{(0)}\left(\omega, \omega^{\prime} ; \mu\right) \phi_{B}^{-}\left(\omega^{\prime} ; \mu\right) +\mathcal{O}\left(\alpha_{s}^{2}\right),
    \notag \\
    \gamma_{+}^{(0)}\left(\omega, \omega^{\prime} ; \mu\right)
    &=
    \left(\Gamma_{\mathrm{c}}^{(0)} \ln \frac{\mu}{\omega}-2\right) \delta\left(\omega-\omega^{\prime}\right)-\Gamma_{\mathrm{c}}^{(0)} \omega\left[\frac{\theta\left(\omega^{\prime}-\omega\right)}{\omega^{\prime}\left(\omega^{\prime}-\omega\right)}+\frac{\theta\left(\omega-\omega^{\prime}\right)}{\omega\left(\omega-\omega^{\prime}\right)}\right]_{+}.
\end{align}
We substitute Eqs.~\eqref{appeq:phip2} and \eqref{appeq:phim2} into Eq.~\eqref{appWWd} and verify that the W-W relation still holds at the two-loop level in the large-$\beta_0$ limit.

\section{Explicit expressions for \texorpdfstring{$J_{n,\bar{n}}^{\pm}$}{J} and spectral representations}
\label{app:2loopresults}
We extract the jet function $J$ from the finite part of the $\Pi_\mu^{hc}$:
    \begin{align}
        J^{(+)}_n=&-\frac{\alpha_s}{4\pi} C_F \frac{\bar{\eta}}{\eta}L_{\bar{\eta}}
  +
  (\frac{\alpha_s}{4\pi})^2 C_F N_f \frac{\bar{\eta}}{\eta}\left(\frac{1}{3}L_{p}^2-\frac{1}{3}L_{\omega}^2+ \frac{19}{9}L_{\bar{\eta}}+ \frac{2}{3}{\rm Li}_2(\eta) \right),\\
        J^{(-)}_n=&1+\mathcal{O}(\alpha_s^2\beta_0),\\
        J^{(+)}_{\bar{n}}=&\frac{n\cdot p}{m_b} J^{(+)}_n,\\
        J^{(-)}_{\bar{n}}=&1+
  (\frac{\alpha_s}{4\pi})C_F\left[
  2L_\omega^2-L_p^2-(1+\frac{2}{\eta})L_{\bar{\eta}}-1-\frac{\pi^2}{6}
  \right] \notag\\
  &
  +(\frac{\alpha_s}{4\pi})^2 C_F N_f \left\{\frac{2}{9}L_{p}^3+\left(\frac{2}{3\eta}+\frac{13}{9}\right)L_{p}^2-\frac{4}{9} L_{\omega}^3-\left(\frac{2}{3\eta}+\frac{23}{9}\right)L_{\omega}^2-\frac{38}{27}L_{\omega} -\frac{2}{3}L_{\eta}\right.\notag\\
&\,\,+\left(\frac{38}{9\eta}+\frac{6\pi^2+95}{27}\right)L_{\bar{\eta}}+\frac{4}{3}\left[\left(L_{\omega}+\frac{1}{\eta}+\frac{4}{3}\right){\rm Li}_2(\eta)-\frac{1}{2}{\rm Li}_3(\eta)\right]\notag\\
&\,\,\left.+\frac{16}{9}\zeta_3+\frac{15\pi^2+14}{162}\right\}.      \end{align}
By computing the imaginary parts of the convolution integrals $J^{\pm} \otimes \phi_B^{\pm}$, we determine the $\mathcal{O}(\alpha_s^2 \beta_0)$ contribution to the $B \to \pi$ form factors in Eq.~\eqref{eq:NNLLsumrule}. We summarize the corresponding spectral functions below. The expressions entering the one-loop correlation functions were derived in Refs.~\cite{Wang:2015vgv,DeFazio:2007hw}. We have confirmed their results and present the missing pieces that arise in the $\mathcal{O}(\alpha_s^2 \beta_0)$ correlation functions.
\begin{align}
    \frac{1}{\pi}&~{\rm Im}_{\omega'}\int_0^\infty \frac{d\omega}{\omega-\omega'-i0}\ln{\frac{\mu^2}{n\cdot p\left(\omega-\omega'-i0\right)}}~\phi_B^-(\omega,\mu)\notag\\
    &=\int_0^\infty d\omega\left[\frac{\theta(\omega'-\omega)}{\omega-\omega'}\right]_\oplus\phi_B^-(\omega,\mu)+\ln{\frac{\mu^2}{n\cdot p~\omega'}}~\phi_B^-(\omega',\mu),\\
    \frac{1}{\pi}&~{\rm Im}_{\omega'}\int_0^\infty \frac{d\omega}{\omega-\omega'-i0}\ln^2{\frac{\mu^2}{n\cdot p\left(\omega-\omega'-i0\right)}}~\phi_B^\pm(\omega,\mu)\notag\\
    &=\int_0^\infty d\omega \left[\frac{2\theta(\omega'-\omega)}{\omega-\omega'}\ln{\frac{\mu^2}{n\cdot p(\omega'-\omega)}}\right]_\oplus\phi_B^\pm(\omega,\mu)+\left[\ln^2{\frac{\mu^2}{n\cdot p~\omega'}}-\frac{\pi^2}{3}\right]\phi_B^\pm(\omega',\mu),\\
    \frac{1}{\pi}&~{\rm Im}_{\omega'}\int_0^\infty \frac{d\omega}{\omega-\omega'-i0}\ln^3{\frac{\mu^2}{n\cdot p\left(\omega-\omega'-i0\right)}}~\phi_B^-(\omega,\mu)\notag\\
    &=\int_0^\infty d\omega \left[\frac{3\theta(\omega'-\omega)}{\omega-\omega'}\left(\ln^2{\frac{\mu^2}{n\cdot p(\omega'-\omega)}}-\frac{\pi^2}{3}\right)\right]_\oplus\phi_B^-(\omega,\mu)\notag\\
    &~~~~+\left[\ln^3{\frac{\mu^2}{n\cdot p~\omega'}}-\pi^2~\ln{\frac{\mu^2}{n\cdot p~\omega'}}\right]\phi_B^-(\omega',\mu),\\
    \frac{1}{\pi}&~{\rm Im}_{\omega'}\int_0^\infty \frac{d\omega}{\omega-\omega'-i0}\ln{\frac{\omega'-\omega+i0}{\omega'}}~\phi_B^-(\omega,\mu)=-\theta(\omega')\int_{\omega'}^\infty d\omega\left[\ln{\frac{\omega-\omega'}{\omega'}}\right]\frac{d}{d\omega}\phi_B^-(\omega,\mu),\\
    \frac{1}{\pi}&~{\rm Im}_{\omega'}\int_0^\infty \frac{d\omega}{\omega-\omega'-i0}\ln^2{\frac{\omega'-\omega+i0}{\omega'}}~\phi_B^-(\omega,\mu)=-\theta(\omega')\int_{\omega'}^\infty d\omega\left[\ln^2{\frac{\omega-\omega'}{\omega'}}-\frac{\pi^2}{3}\right]\frac{d}{d\omega}\phi_B^-(\omega,\mu),\\
    \frac{1}{\pi}&~{\rm Im}_{\omega'}\int_0^\infty \frac{d\omega}{\omega-\omega'-i0}\ln^3{\frac{\omega'-\omega+i0}{\omega'}}~\phi_B^-(\omega,\mu)\notag\\
    &=-\theta(\omega')\int_{\omega'}^\infty d\omega\left[\ln^3{\frac{\omega-\omega'}{\omega'}}-\pi^2~\ln{\frac{\omega-\omega'}{\omega'}}\right]\frac{d}{d\omega}\phi_B^-(\omega,\mu),\\
    \frac{1}{\pi}&~{\rm Im}_{\omega'}\int_0^\infty \frac{d\omega}{\omega}\ln{\frac{\omega'-\omega+i0}{\omega'}}~\phi_B^-(\omega,\mu)=\theta(\omega')\int_{\omega'}^\infty\frac{d\omega}{\omega}\phi_B^-(\omega,\mu),\\
    \frac{1}{\pi}&~{\rm Im}_{\omega'}\int_0^\infty \frac{d\omega}{\omega}\ln^2{\frac{\omega'-\omega+i0}{\omega'}}~\phi_B^-(\omega,\mu)=\theta(\omega')\int_{\omega'}^\infty\frac{d\omega}{\omega}\left[2\ln{\frac{\omega-\omega'}{\omega'}}\right]\phi_B^-(\omega,\mu),\\
    \frac{1}{\pi}&~{\rm Im}_{\omega'}\int_0^\infty \frac{d\omega}{\omega}\ln^3{\frac{\omega'-\omega+i0}{\omega'}}~\phi_B^-(\omega,\mu)=\theta(\omega')\int_{\omega'}^\infty\frac{d\omega}{\omega}\left[3\ln^2{\frac{\omega-\omega'}{\omega'}}-\pi^2\right]\phi_B^-(\omega,\mu),\\
    \frac{1}{\pi}&~{\rm Im}_{\omega'}\int_0^\infty \frac{d\omega}{\omega-\omega'-i0}\ln{\frac{\mu^2}{n\cdot p~(\omega-\omega'-i0)}}\ln{\frac{\omega'-\omega+i0}{\omega'}}~\phi_B^-(\omega,\mu)\notag\\
    &=\int_0^\infty d\omega\left[\frac{\theta(\omega'-\omega)}{\omega-\omega'}\ln{\frac{\omega'-\omega}{\omega'}}\right]_\oplus \phi_B^-(\omega,\mu)\notag\\
    &~~~~+\frac{1}{2}\int_{\omega'}^\infty d\omega \left[\ln^2{\frac{\mu^2}{n\cdot p~(\omega-\omega')}}-\ln^2{\frac{\mu^2}{n\cdot p~\omega'}}+\frac{\pi^2}{3}\right]\frac{d}{d\omega}{\phi_B^-(\omega,\mu)},\\
    \frac{1}{\pi}&~{\rm Im}_{\omega'}\int_0^\infty \frac{d\omega}{\omega-\omega'-i0}\ln{\frac{\mu^2}{n\cdot p~(\omega-\omega'-i0)}}\ln^2{\frac{\omega'-\omega+i0}{\omega'}}~\phi_B^-(\omega,\mu)\notag\\
    &=
    {
    -\frac{1}{3}\left[\mathcal{P}\int_0^\infty d\omega~\ln^3{\frac{\omega'-\omega}{\omega'}}
    \frac{d}{d\omega}\phi_B^-(\omega,\mu)
    +
    \pi^2\ln{\frac{\mu^2}{n\cdot p~\omega'}}~\phi_B^-(\omega',\mu)\right]
    }
    \notag\\
    &~~~~-\int_{\omega'}^\infty d\omega~\ln^2{\frac{\omega-\omega'}{\omega'}}\ln{\frac{\mu^2}{n\cdot p~(\omega-\omega')}}\frac{d}{d\omega}\phi_B^-(\omega,\mu)
    \\
    \frac{1}{\pi}&~{\rm Im}_{\omega'}\int_0^\infty \frac{d\omega}{\omega}\ln{\frac{\mu^2}{n\cdot p~(\omega-\omega'-i0)}}\ln{\frac{\omega'-\omega+i0}{\omega'}}~\phi_B^-(\omega,\mu)\notag\\
    &=\int_{\omega'}^\infty \frac{d\omega}{\omega}~\ln{\frac{\mu^2}{n\cdot p~(\omega-\omega')}}~\phi_B^-(\omega,\mu)+\int_0^{\omega'}\frac{d\omega}{\omega}~\ln{\frac{\omega'-\omega}{\omega'}}\phi_B^-(\omega,\mu)\\
    \frac{1}{\pi}&~{\rm Im}_{\omega'}\int_0^\infty \frac{d\omega}{\omega-\omega'-i0}~{\rm Li}_2\left(\frac{\omega}{\omega'}\right)\phi_B^-(\omega,\mu)=\frac{\pi^2}{6}~\phi_B^-(\omega',\mu)-\int_{\omega'}^\infty \frac{d\omega}{\omega-\omega'}\ln{\frac{\omega}{\omega'}}~\phi_B^-(\omega,\mu),\\
    \frac{1}{\pi}&~{\rm Im}_{\omega'}\int_0^\infty \frac{d\omega}{\omega-\omega'-i0}~\frac{\omega'}{\omega}~{\rm Li}_2\left(\frac{\omega}{\omega'}\right)\phi_B^\pm(\omega,\mu)\notag\\
    &=\frac{\pi^2}{6}~\phi_B^\pm(\omega',\mu)-\int_{\omega'}^\infty \frac{d\omega}{\omega-\omega'}~\frac{\omega'}{\omega}\ln{\frac{\omega}{\omega'}}~\phi_B^\pm(\omega,\mu),\\
    \frac{1}{\pi}&~{\rm Im}_{\omega'}\int_0^\infty \frac{d\omega}{\omega-\omega'-i0}~{\rm Li}_3\left(\frac{\omega}{\omega'}\right)\phi_B^\pm(\omega,\mu)=\zeta(3)~
    {\phi_B^\pm(\omega',\mu)}
    -\frac{1}{2}\int_{\omega'}^\infty \frac{d\omega}{\omega-\omega'}\ln^2{\frac{\omega}{\omega'}}~\phi_B^\pm(\omega,\mu),\\
    \frac{1}{\pi}&~{\rm Im}_{\omega'}\int_0^\infty \frac{d\omega}{\omega-\omega'-i0}~{\rm Li}_2\left(\frac{\omega}{\omega'}\right)\ln{\frac{\mu^2}{n\cdot p~(\omega-\omega'-i0)}}~\phi_B^-(\omega,\mu)\notag\\
    &=\int_0^\infty d\omega\left[\frac{\theta(\omega'-\omega)}{\omega-\omega'}\right]_\oplus{\rm Li}_2\left(\frac{\omega}{\omega'}\right)\phi_B^-(\omega,\mu)-\int_{\omega'}^\infty \frac{d\omega}{\omega-\omega'}\ln{\frac{\omega}{\omega'}}\ln{\frac{\mu^2}{n\cdot p~(\omega-\omega')}}\phi_B^-(\omega,\mu)\notag\\
    &~~~~+\frac{\pi^2}{6}\ln{\frac{\mu^2}{n\cdot p~\omega'}}~\phi_B^-(\omega',\mu),\\
    \frac{1}{\pi}&~{\rm Im}_{\omega'}\int_0^\infty \frac{d\omega}{\omega-\omega'-i0}~\left[{\rm Li}_3\left(\frac{\omega'-\omega+i0}{\omega'}\right)+\frac{1}{2}\ln^2{\frac{\omega'-\omega+i0}{\omega'}}\ln{\frac{\omega}{\omega'}}\right]~\phi_B^-(\omega,\mu)\notag\\
    &=\int_{\omega'}^\infty \frac{d\omega}{\omega-\omega'}\ln{\frac{\omega}{\omega'}}\ln{\frac{\omega-\omega'}{\omega'}}~\phi_B^-(\omega,\mu),\\
    \frac{1}{\pi}&~{\rm Im}_{\omega'}\int_0^\infty \frac{d\omega}{\omega-\omega'-i0}\ln^2{\frac{\mu^2}{-n\cdot p~(\omega'+i0)}}~\phi_B^\pm(\omega,\mu)\notag\\
    &=\mathcal{P}\int_0^\infty\frac{d\omega}{\omega-\omega'}\left(2\ln{\frac{\mu^2}{n\cdot p~ \omega'}}\right)~\phi_B^\pm(\omega,\mu)+\left(\ln^2{\frac{\mu^2}{n\cdot p~\omega'}}-\pi^2\right)\phi_B^\pm(\omega',\mu),\\
    \frac{1}{\pi}&~{\rm Im}_{\omega'}\int_0^\infty \frac{d\omega}{\omega-\omega'-i0}\ln^3{\frac{\mu^2}{-n\cdot p~(\omega'+i0)}}~\phi_B^-(\omega,\mu)\notag\\
    &=\mathcal{P}\int_0^\infty\frac{d\omega}{\omega-\omega'}\left(3\ln^2{\frac{\mu^2}{n\cdot p~ \omega'}}-\pi^2\right)~\phi_B^-(\omega,\mu)+\left(\ln^3{\frac{\mu^2}{n\cdot p~\omega'}}-3\pi^2\ln{\frac{\mu^2}{n\cdot p~\omega'}}\right)\phi_B^-(\omega',\mu),
\end{align}

with the $\oplus$-function defined by
\begin{align}
    \int_{0}^{\infty} d \omega\left[f\left(\omega', \omega\right)\right]_{\oplus} g\left(\omega\right)=\int_{0}^{\infty} d \omega f\left(\omega', \omega\right)\left[g\left(\omega\right)-g(\omega')\right].
\end{align}
Substituting the above integrals into Eqs.(C.1)-(C.4), we present the explicit results for the effective distribution amplitudes $\Phi_{n,\bar n}^{\pm}(\omega')$
\begin{align}
\boldsymbol{\Phi}_{n}^{+}(\omega^{\prime},\mu)
&=\frac{\alpha_s C_F}{4 \pi}\int_{\omega^{\prime}}^{\infty} \frac{d \omega}{\omega} \phi_B^{+}(\omega,\mu) + \left(\frac{\alpha_s}{4 \pi}\right)^2 C_F \beta_0 \left[ \int_0^{\omega^{\prime}} \frac{d \omega}{\omega} \ln \frac{\omega^{\prime}-\omega}{\omega^{\prime}} \phi_B^{+}(\omega,\mu) \right.  \notag\\
&  \left. - \int_{\omega^{\prime}}^{\infty} \frac{d \omega}{\omega} \left(\ln \frac{\omega}{\omega^{\prime}} - \ln \frac{\mu^2}{n \cdot p(\omega-\omega^{\prime})} - \ln \frac{\omega-\omega^{\prime}}{\omega^{\prime}} - \frac{19}{6} \right) \phi_B^{+}(\omega,\mu) \right] ,\\
\boldsymbol{\Phi}_{n}^{-}(\omega^\prime,\mu)&=\phi_B^{-}(\omega',\mu),\\
\boldsymbol{\Phi}_{\bar{n}}^{+}(\omega^{\prime},\mu)
&=  \frac{n \cdot p}{{m_b}} \boldsymbol{\Phi}_{n}^{+}(\omega^\prime,\mu) ,\\
\boldsymbol{\Phi}_{\bar{n}}^{-}(\omega^{\prime},\mu)
&= \phi_B^-(\omega',\mu) + \frac{\alpha_s C_F}{4\pi} \left\{ \int_0^{\omega'} d\omega \left[ \frac{2}{\omega - \omega'} \left( \ln \frac{\mu^2}{n \cdot p (\omega' - \omega)} - \ln \frac{\omega' - \omega}{\omega'} \right) \right]_{\oplus} \phi_B^-(\omega,\mu) \right. \notag\\
&- \left.\int_{\omega'}^\infty d\omega \left[ \ln^2 \frac{\mu^2}{n \cdot p \omega'} - \left( 2 \ln \frac{\mu^2}{n \cdot p \omega'} + 3 \right) \ln \frac{\omega - \omega'}{\omega'} + 2 \ln \frac{\omega}{\omega'} + \frac{\pi^2}{6} - 1 \right] \frac{d \phi_B^-(\omega,\mu)}{d\omega}\right\} \notag\\
&+ \left( \frac{\alpha_s}{4\pi} \right)^2 C_F \beta_0 \left\{ -\int_0^{\omega'} \left[ \frac{d\omega}{\omega - \omega'} \right]_{\oplus} \left[ {2} \text{Li}_2 \left( \frac{\omega}{\omega'} \right) \phi_B^-(\omega,\mu) \right] \right. \notag\\
&+\int_0^{\omega'} \left[ \frac{d\omega}{\omega - \omega'} \left( 2 \ln^2 \frac{\mu^2}{n \cdot p (\omega' - \omega)} + \frac{29}{3} \ln \frac{\mu^2}{n \cdot p (\omega' - \omega)} + \frac{19}{9} - \frac{2\pi^2}{3} \right) \right]_{\oplus} \phi_B^-(\omega,\mu) \notag \\
&- \int_0^{\omega'} d\omega \left\{ \left[ \frac{1}{\omega - \omega'}\left( \ln^2 \frac{\mu^2}{n \cdot p \omega'} + \frac{19}{3} \ln \frac{\mu^2}{n \cdot p \omega'} - {\frac{\pi^2}{3}} \right) \right]_{{\oplus}}  {-}\frac{2}{{\omega}} \ln \frac{\omega' - \omega}{\omega'} \right\} \phi_B^-(\omega,\mu) \notag\\
&+ \int_{\omega'}^\infty \frac{d\omega}{\omega - \omega'} \left[ 2 \ln \frac{\mu^2}{n \cdot p \omega'} \left( \ln \frac{\omega}{\omega'} + \frac{\omega - \omega'}{\omega} \right) + \frac{19}{3} \frac{\omega - \omega'}{\omega} - \frac{1}{2} \ln^2 \frac{\omega}{\omega'} +  \frac{2\omega'}{\omega} \ln \frac{\omega}{\omega'} \right.\notag \\
&\left. ~~~~+ \frac{8}{3} \ln \frac{\omega}{\omega'} \right] \phi_B^-(\omega,\mu) \notag\\
&+ \int_{\omega'}^\infty d\omega \left[ {-}\frac{1}{3} \ln^3 \frac{\mu^2}{n \cdot p \omega'} {-}\frac{5}{3} \ln^2 \frac{\mu^2}{n \cdot p \omega'} {-}\frac{19}{9} \ln \frac{\mu^2}{n \cdot p \omega'}   \right. \notag\\
&\left. \left. + {\left( \ln^2 \frac{\mu^2}{n \cdot p \omega'} + \frac{19}{3} \ln \frac{\mu^2}{n \cdot p \omega'} + \frac{209}{18} \right)} \ln \frac{\omega - \omega'}{\omega'}{-}\frac{23\pi^2}{36} {+}\frac{7}{54} + {\frac{5\zeta(3)}{3}} \right] \frac{d\phi_B^-(\omega,\mu)}{d\omega} \right\}.
\end{align}

\section{Modeling the \texorpdfstring{$B$}{B}-meson LCDAs}\label{app-B-LCDA}
The three-parameter model for leading-twist and higher-twist $B$-meson LCDAs at the reference scale $\mu_0$ can be constructed in Ref.\cite{Beneke:2018wjp,Wang:2021yrr,Gao:2021sav}.
\begin{align}
\phi_B^+(\omega) &= \omega \, \mathbb{F}(\omega;-1) \,, \qquad\quad
\phi_B^{-\rm WW}(\omega) = \mathbb{F}(\omega;0) \,,
\nonumber \\
\phi_B^{-\rm t3}(\omega) &= \frac{1}{6}\,\mathcal{N} \,
(\lambda^2_E - \lambda^2_H ) \,
\Big[
-\omega^2\,\mathbb{F}(\omega;-2)
+ 4\,\omega\,\mathbb{F}(\omega;-1) -2\, \mathbb{F}(\omega;0)
\Big] \,,
\nonumber \\
\phi_3(\omega_1,\omega_2) &=  \frac{1}{2}\,\mathcal{N} \,
(\lambda^2_E - \lambda^2_H ) \,
\omega_1\,\omega^2_2\, \mathbb{F}(\omega_1+\omega_2;-2) \,,
\nonumber \\
\hat{g}_B^+(\omega) &=
\frac{1}{4}\bigg[
2\,\omega\,(\omega-\bar{\Lambda})\,\mathbb{F}(\omega;0)
+(3\omega-2\bar{\Lambda})\,\mathbb{F}(\omega;1)
+3\,\mathbb{F}(\omega;2)
\nonumber \\
& -\frac{1}{6} \, \mathcal{N} \, (\lambda^2_E - \lambda^2_H ) \,
\omega^2 \,\mathbb{F}(\omega;0)\bigg]\,,
\nonumber \\
\hat{g}_B^-(\omega) &=
\frac{1}{4}\,\bigg\{ (3\omega-2\bar{\Lambda})\,\mathbb{F}(\omega;1)
+3\, \mathbb{F}(\omega;2)
\nonumber \\
& +\frac{1}{3} \, \mathcal{N} \,
(\lambda^2_E - \lambda^2_H ) \,
\omega\,\bigg[
\omega\, (\bar{\Lambda}-\omega) \, \mathbb{F}(\omega;-1)
-\Big(2\,\bar{\Lambda} -\frac{3}{2}\,\omega \Big) \,
\mathbb{F}(\omega;0) \bigg]\bigg\}\,,
\nonumber \\
\phi_4(\omega_1,\omega_2) &= \frac{1}{2}\,\mathcal{N} \,
(\lambda^2_E + \lambda^2_H ) \,
\omega^2_2\, \mathbb{F}(\omega_1+\omega_2;-1) \,,
\nonumber \\
\psi_4(\omega_1,\omega_2)
&= \mathcal{N} \,
\lambda^2_E  \,
\omega_1\, \omega_2\,  \mathbb{F}(\omega_1+\omega_2;-1) \,,\qquad
\tilde{\psi}_4(\omega_1,\omega_2) = \mathcal{N} \,
\lambda^2_H  \,
\omega_1\, \omega_2\,  \mathbb{F}(\omega_1+\omega_2;-1) \,,
\nonumber \\
\phi_5(\omega_1,\omega_2) &= \mathcal{N} \,
(\lambda^2_E + \lambda^2_H ) \,
\omega_1\, \mathbb{F}(\omega_1+\omega_2;0) \, , \hspace{2.7mm}
\psi_5(\omega_1,\omega_2) = - \mathcal{N} \,
\lambda^2_E \,
\omega_2\,  \mathbb{F}(\omega_1+\omega_2;0) \,,
\nonumber \\
\tilde{\psi}_5(\omega_1,\omega_2) &= - \mathcal{N} \,
\lambda^2_H \,
\omega_2\,  \mathbb{F}(\omega_1+\omega_2;0) \,,
\nonumber \\
\phi_6(\omega_1,\omega_2) &= \mathcal{N} \,
(\lambda^2_E - \lambda^2_H ) \, \mathbb{F}(\omega_1+\omega_2;1)\,,\label{model:ansatz}
\end{align}
where
\begin{align}
\mathcal{N} &= \frac{1}{3}\,\frac{\beta\,(\beta+1)}{\alpha\,(\alpha+1)}\,
\frac{1}{\omega^2_0} \, ,\qquad\qquad\qquad
\bar{\Lambda} = \frac{3}{2}\, \frac{\alpha}{\beta} \, \omega_0\,,
\notag\\
\mathbb{F}(\omega;n) &\equiv
\omega^{n-1}_0 \,
U \left ( \beta-\alpha,2-n-\alpha,{\frac{\omega}{\omega_0}} \right ) \,
\frac{\Gamma(\beta)}{\Gamma(\alpha)}\,
e^{-\frac{\omega}{\omega_0}} \,,
\label{model:bb-F}
\end{align}
with $U(a, b, z)$ being the hypergeometric $U$ function.

The evolution of the two-particle DAs can be written as a compact analytic function,
\begin{align}
\phi_B^+(\omega,\mu)
=&~  U_\phi(\mu,\mu_0) \,  \frac{1}{\omega^{p+1}}\,
\frac{\Gamma(\beta)}{\Gamma(\alpha)} \,
\mathcal{G}(\omega;0,2,1)\,,
\nonumber \\
\phi^{-\rm WW}_B(\omega,\mu)
=&~   U_\phi(\mu,\mu_0) \,  \frac{1}{\omega^{p+1}}\,
\frac{\Gamma(\beta)}{\Gamma(\alpha)} \,
\mathcal{G}(\omega;0,1,1)\,,
\nonumber \\
\phi^{-\rm t3}_B(\omega,\mu)
=&  - \frac{1}{6}\, U^{\rm t3}_\phi(\mu,\mu_0) \,\mathcal{N} \,
(\lambda^2_E - \lambda^2_H ) \,\frac{\omega^2_0}{\omega^{p+3}} \,
\frac{\Gamma(\beta)}{\Gamma(\alpha)}\,
\bigg\{ \mathcal{G}(\omega;0,3,3)
\nonumber \\
&
+ (\beta-\alpha) \, \bigg[ \frac{\omega}{\omega_0}\,
\mathcal{G}(\omega;0,2,2)
-\beta \, \frac{\omega}{\omega_0}\,
\mathcal{G}(\omega;1,2,2)
- \mathcal{G}(\omega;1,3,3)\bigg]
\bigg\} \,,
\label{eq:LL_lcda}
\end{align}
where $\displaystyle p= \left ( \Gamma^{(0)}_{\rm cusp}/2\beta_0 \right ) \ln[\alpha_{s}(\mu)/\alpha_{s}(\mu_{0})]$, and the explicit expressions for Meijer $\mathcal{G}$-function and evolution function $U$ can be found in Refs.\cite{Beneke:2018wjp,Gao:2021sav}.

    \bibliography{main}
    \bibliographystyle{JHEP}

\end{document}